\documentclass{article}
\usepackage[a4paper,total={6.5in,8.5in}]{geometry}
\usepackage{macros}
\usepackage{subcaption}
\usepackage{graphicx}
\usepackage{algorithm}
\usepackage{algpseudocode}
\usepackage{booktabs} 
\usepackage{multirow} 
\usepackage[numbers, sort, comma, square]{natbib}
\usepackage{authblk}
\usepackage{wrapfig}
\usepackage{amsmath}
\usepackage{amssymb}
\usepackage{colortbl}
\usepackage{tcolorbox}
\usepackage{appendix,minitoc}
\usepackage{titletoc}

\newcommand\DoToC{%
  \startcontents
\hypersetup{colorlinks=true, linkcolor=pierCite}
  \printcontents{}{1}{\subsection*{\textbf{Table of contents}}}
  \vskip3pt\vskip5pt
}

\graphicspath{{figures/}}
\author{Piersilvio De Bartolomeis}
\author{Javier Abad}
\author{Konstantin Donhauser}
\author{Fanny Yang}
\affil{Department of Computer Science, ETH Zürich}

\title{Detecting critical treatment effect bias in small subgroups}

\date{}
\begin{document}
\maketitle 

\begin{abstract}
Randomized trials are considered the gold standard for making informed decisions in medicine, yet they often lack generalizability to the patient populations in clinical practice.  Observational studies, on the other hand, cover a broader patient population but are prone to various biases. Thus, before using an observational study for decision-making, it is crucial to \emph{benchmark} its treatment effect estimates against those derived from a randomized trial. 
We propose a novel strategy to benchmark observational studies beyond the average treatment effect. First, we design a statistical test for the null hypothesis that the treatment effects estimated from the two studies, conditioned on a set of relevant features, differ up to some tolerance. We then estimate an asymptotically valid lower bound on the maximum bias strength for any subgroup in the observational study.  Finally, we validate our benchmarking strategy in a real-world setting and show that it leads to conclusions that align with established medical knowledge\footnote{See our GitHub repository for the source code: \url{https://github.com/jaabmar/kernel-test-bias}.}.
\end{abstract}

\section{Introduction}

Randomized trials have traditionally been the gold standard for informed decision-making in medicine, as they allow for unbiased estimation of treatment effects under mild assumptions. However, there is often a significant discrepancy between the patients observed in clinical practice and those enrolled in randomized trials, limiting the generalizability of the trial results~\citep {rothwell2005external, duma2018representation}. To address this issue, the U.S. Food and Drug Administration advocates for using observational data, as it is usually more representative of the patient population in clinical practice~\citep{platt2018fda, klonoff2020new}. Yet, a major caveat to this recommendation is that several sources of bias, including hidden confounding, can
compromise the causal conclusions drawn from observational data. 

In light of the inherent limitations of  randomized and observational data, it has become a popular strategy to \emph{benchmark} observational studies against existing randomized trials to assess their quality \citep{dahabreh2020benchmarking, forbes2020benchmarking}. The main idea behind this approach is first to emulate the procedures adopted in the randomized trial within the observational study; see e.g.~\citep{hernan2016using} for a detailed explanation. 
Then, the treatment effect estimates from the observational data are compared with those from the randomized data. If the estimates are similar, we may be willing to trust the observational study for patient populations where the randomized data is insufficient.

To support the benchmarking framework, several works propose statistical tests that compare treatment effect estimates between randomized and observational data~\citep{viele2014use,hussain2023falsification,de2023hidden,yangelastic,demirel2024benchmarking}. In particular, two properties have been identified as essential for effective benchmarking of observational studies: \emph{tolerance} and \emph{granularity}.  
Tolerance allows the acceptance of studies with negligible bias that does not impact decision-making, thereby significantly reducing false rejections in real-world settings where some bias is expected. Granularity, on the other hand, allows the detection of bias on small subgroups or individuals that would otherwise go unnoticed.  

However, to date, no existing statistical test satisfies both properties. Our contributions here are as follows.
\begin{itemize}
	\item We design a statistical test for the null hypothesis that treatment effects estimated from the two studies, conditioned on a set of features that define the patient subgroups, differ up to some tolerance. To our knowledge, our test is the first to satisfy tolerance and granularity. We then leverage both properties to estimate an asymptotically valid lower bound on the maximum bias in the observational study. 
	\item  We propose a novel strategy to benchmark observational studies. Specifically, we compare the lower bound on the bias against a \emph{critical value}, e.g. the minimum bias strength that would explain away the estimated treatment effect in a subgroup of interest. If the lower bound is greater than the critical value, we discard the conclusions drawn from the observational study. Finally, we demonstrate that our strategy yields conclusions consistent with current epidemiological knowledge using real-world data.
\end{itemize}

\section{Problem setting}
\label{sec:setting}
We have access to two datasets: $\datarct$ of size $\nrct$  from a randomized trial~($\rct$) and  $\dataobs$ of size $\nobs$ from an observational study~($\obs$), containing tuples $Z := (X,Y,T)$ of covariates $X \in \RR^\xdim$,  bounded observed outcome $Y\in \RR$, and treatment assignment variable $T \in \{0,1\}$. We assume that the data is drawn i.i.d~from the distributions $\prct$ and $\pobs$
that are marginal distributions of the respective 
full distribution $\pfull^\diamond$
over $\left(X, U, Y(0), Y(1),Y, T\right)$ for $\diamond \in \{\rct, \obs\}$. In particular, the full distribution also includes randomness over a vector of unobserved covariates  $U\in \mathbb R^\udim$ and potential outcomes $\left(Y(0), Y(1)\right) \in \RR^2$. We further assume that 
 $\datarct$ and $\dataobs$ are independent of each other, and 
the support of the randomized trial is included in the support of the observational study, i.e.
$\supp(\prct_X) \subseteq \supp(\pobs_X)$, where we use the shorthand $\mathbb P_X$ to denote the marginal distribution of $X$ under $\mathbb P$. 

\paragraph{Treatment effect estimation}
\label{sec:hte}
A crucial quantity to estimate for decision-making in many domains is the conditional average treatment effect~(CATE). The CATE  is a function   $\mu: \XX \to \RR$,  defined by
\begin{align*}
\yone^\diamond (x) \defeq \EE_{\pfull^{\diamond}}\left [  Y(1) - Y(0) \mid X=x \right ]  , \quad \mathrm{for} \;\; \diamond \in \{\rct,\obs\}\;\;\mathrm{and} \;\; \XX \subseteq \supp\left( \pxrct \right).
\end{align*}
Unfortunately, we cannot estimate the CATE from the observed data as we never observe the potential outcomes. Instead, we can estimate the regression function $\estimand:\XX \to \RR$, defined by
\begin{align*}
 \estimand^\diamond(x) \defeq \EE_{\PP^\diamond}\left [ Y \mid T=1, X=x\right ]  -\EE_{\PP^\diamond}\left [ Y \mid T=0, X=x\right ] , \quad \mathrm{for} \;\; \diamond \in \{\rct,\obs\}.
\end{align*}
For the treatment effect in the randomized trial, we can then observe that  $\estimandrct(x) = \caterct(x)$ holds for all $x \in \XX$, under the assumption of internal validity outlined below.
\begin{assumption}[Internal validity]
\label{asm:internalvalid}
The data-generating process of the randomized trial satisfies
 \begin{align*}
(i)&\;\;   Y = Y(T)\;\; \pfullrct -\mathrm{almost\;surely}. \\
(ii)&\;\; T \ind (Y(1),Y(0)).\\
(iii)& \;\; \pfullrct(T =1 \mid X,U)= \pi\in(0,1).
 \end{align*}
\end{assumption} 
In particular, \Cref{asm:internalvalid}
is expected to hold by design in a completely randomized experiment, and thus, $\caterct$ can be estimated from the observed data under mild assumptions~\citep{rubin1978bayesian}. On the other hand, we cannot estimate $\cateobs$ from the observed data due to hidden confounding or other sources of bias in the observational study, i.e. we cannot rule out the existence of $x \in \XX$ such that $\estimandobs (x)\neq \cateobs (x)$. Therefore, it is crucial to benchmark the observational study before using the estimate of $\estimandobs$ for any downstream task.

\begin{figure*}
 \centering 
\includegraphics[scale=0.29]{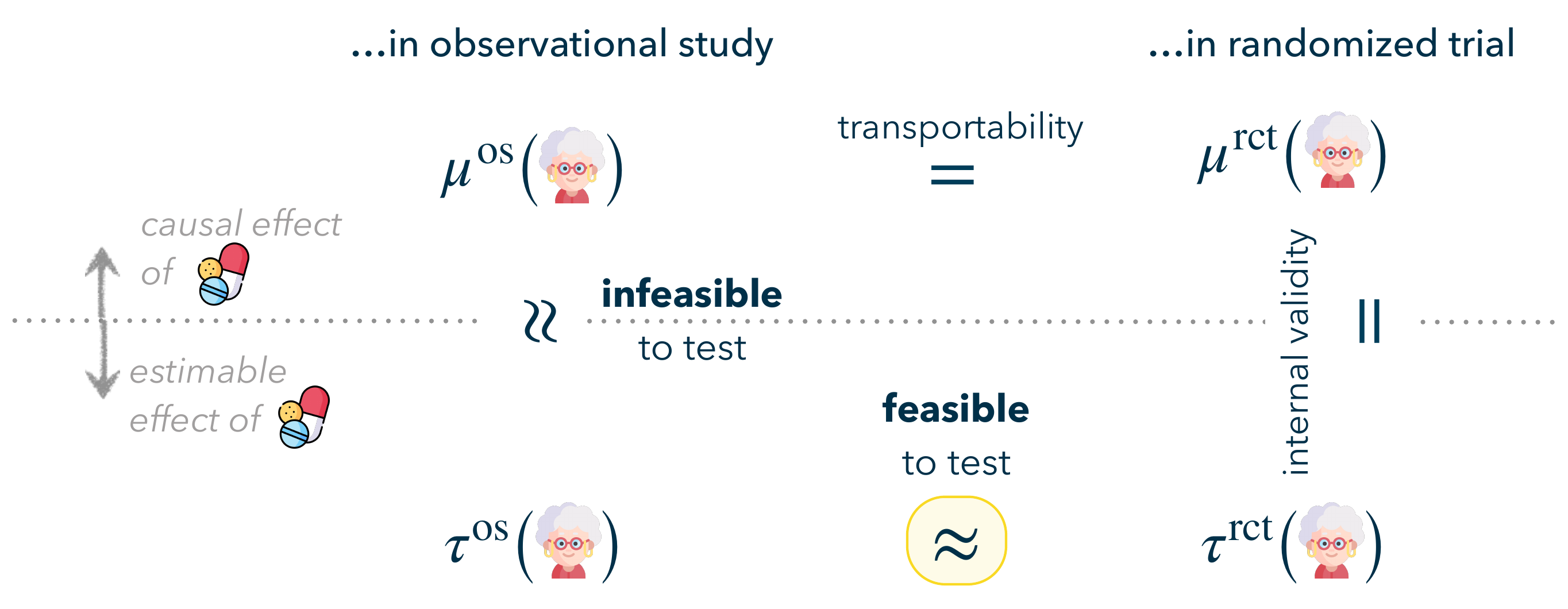} 
    \caption{High-level illustration of our approach. We want to test if the bias in the observational study, i.e. $\cateobs -\estimandobs$, is contained within a tolerance range. However, the true treatment effect $\cateobs$ is not identifiable, and instead, we test the bias between the treatment effects estimated from the two studies, i.e. $ \estimandobs -\estimandrct$.} 
    \label{fig:setting} 
\end{figure*}

\subsection{Null hypothesis}
Our goal is to test if the bias in the observational study, defined as  $ \truedelta(x) \defeq \estimandobs(x) -\cateobs(x)$ for all $x \in \XX$,
is contained within a tolerance range. However, the bias $\truedelta$ is not estimable from the data. Instead, we can test the bias $ \proxybias(x) \defeq \estimandobs(x) - \estimandrct(x)$,  which is equivalent to  $\truebias$ under internal validity and transportability, i.e. $\cateobs(x)=\caterct(x)$ for all $x \in \XX$~(see \Cref{fig:setting}).
 In particular, we would like to test if the bias $\proxybias$ between the two studies is contained within a tolerance range (requires tolerance) across all patient subgroups (requires granularity).  Hence, we will now introduce a null hypothesis that allows for both tolerance and granularity.

To do so, we define two bounded tolerance functions $\estimandobs_{\pm}: \XX \to \RR$ that capture how much the estimated treatment effects can differ between studies and satisfy $ \estimandobs_-(x) \leq \estimandobs(x) \leq  \estimandobs_+(x) $ for all $x \in \XX$.
Further, we define the patient subgroups via a subset of features $\subx$, corresponding to the covariates with indices  $\mathcal J \subseteq \{1, \cdots, d\}$. We can then introduce our null hypothesis,  given by 
\begin{align}
\label{eq:catetolnull}
 \hnull: \;\; \pcaterct \in 
    \left[\pcateobslb, \pcateobsub \right], \;\;\prct_{\subx}-\mathrm{almost\;surely}.\end{align}
   \paragraph{Discussion of our null hypothesis} We provide several remarks on  the null hypothesis in~\Cref{eq:catetolnull}. 
First, we satisfy tolerance by testing if $\estimandrct(x)$ is contained (in probability) in an interval around $\estimandobs(x)$, for all $x \in \XX$. Second, we can satisfy granularity by choosing an appropriate subset $\mathcal J$: When $|\mathcal J| = d$, we detect bias at the individual level, thereby satisfying the strictest definition of granularity. On the other hand, when $|\mathcal J| = 0$, we test if the average treatment effects are equal, thus potentially ignoring bias in small subgroups and individuals. Third,  we test if the treatment effects are equal (up to tolerance) on the support of the randomized trial since we cannot extrapolate outside the support of $\pxrct$ without further assumptions. 
\paragraph{Example 1: User-specified tolerance}
\label{sec:transportability}
A natural choice for the tolerance functions is to add (respectively subtract) a user-specified function $\delta(x) \geq 0$,  that is
\begin{align*}
\estimandobs_{\pm}(x) = \estimandobs (x) \pm \delta(x), \quad \mathrm{for\;all}\; x \in \XX.
\end{align*}
The function $\delta$ can incorporate all sources of bias in the observational study, such as unobserved confounding and non-adherence to treatment assignments. For instance, we can test whether $\|\proxybias\|_{L^\infty(\pxrct)}$ is larger than a critical value $\deltactoracle \in \RR$ by choosing $\estimandobs_{\pm}(x) = \estimandobs (x) \pm \deltactoracle$. 
Further, similar tolerance functions have been previously used in the context of modeling violations of the transportability assumption, see e.g. \citep{nguyen2017sensitivity,nguyen2018sensitivity, dahabreh2022global,dahabreh2023sensitivity}.

\paragraph{Example 2: Sensitivity analysis bounds}
Another practical choice for the tolerance functions $\estimandobs_{\pm}$ is to use the upper and lower bounds arising from a sensitivity analysis model. For instance, the marginal sensitivity model~\citep{tan2006distributional} is commonly used to account for unobserved confounding in observational data. In particular, this model assumes that the influence of $U$ on $T$ is limited by a \emph{confounding strength} $\Gamma$, i.e.
\begin{align}
\label{eq:msm}
\confvalue^{-1} \leq \frac{\pfullobs(T=1|X,U)}{\pfullobs(T=0|X,U)} / \frac{\pobs(T=1|X)}{\pobs(T=0|X)} \leq \confvalue, \quad \pfullobs-\mathrm{almost\;surely.}
\end{align}
We can thus define $\estimandobs_{\pm}$ as the upper and lower bounds for $\estimandobs$ under the assumption of $\confvalue$-bounded confounding strength. Our test can then be used to detect if the marginal sensitivity model is well-specified, i.e. if~\Cref{eq:msm} holds for a specific choice of $\Gamma$; see e.g.~\citep {de2023hidden} for a more detailed explanation of this setting.

\section{Methodology}
\label{sec:test}
 In this section, we rewrite the null hypothesis from \Cref{eq:catetolnull} in terms of a \emph{signal} function that captures the bias between $\estimandobs$ and $\estimandrct$.  Then, we propose an oracle test statistic assuming that the tolerance functions $\estimandobs_\pm$ are known. Finally, we provide asymptotic guarantees for the finite-sample test statistic where the tolerance functions are estimated from the observational data.

\subsection{Null hypothesis using signal function}
We first observe that, for some tolerance functions $\estimandobs_\pm$, \Cref{eq:catetolnull} is equivalent to stating that there exists a function $g: \RR^{|\subsetx|} \to [0,1]$ such that $\estimandobs_g(X) \defeq g \left(\subx \right)\ubobs(X) + \left(1-g \left(\subx \right)\right)\lbobs(X)$ satisfies
$$
\pcaterct
    =   \pcateobsg, \quad \prct_{\subx}-\mathrm{almost\;surely}.
$$
We test a slightly more restrictive hypothesis by assuming that $g$ lies in a sufficiently rich function class $\mathcal G$:
\begin{align*}
    \hnull^\GG: \pcaterct
    =   \pcateobstrueg, \quad \mathrm{for\;some}\; \trueg \in \GG,\quad \prct_{\subx}-\mathrm{almost\;surely}.
 \end{align*}  
In practice, one can either restrict $\GG$ to a particular function class if domain knowledge is available or use neural networks as general function approximations, for which the assumption is expected to hold.

We can then rewrite the null hypothesis above using a \emph{signal} function that captures the bias between the estimates from observational and randomized data. 
Indeed, for any choice of $\subsetx$, we have by~\Cref{asm:internalvalid}
$$
\EE_{\prct} [\estimandrct(X) \mid \subx = x] = \EE_{\prct}\left [  Y \left(\frac{T}{\pi}-\frac{1-T}{1-\pi}\right)   \mid \subx = x \right ],\;\; \mathrm{for\;all}\;x \in \supp\left(\pxrctj\right).
$$
Further, recall that $Z=(X,Y,T)$ is the vector of observed variables, and thus
 by defining the signal function
\begin{align*}
   \psi_{g}(Z) &=  Y \left(\frac{T}{\pi}-\frac{1-T}{1-\pi}\right) - \estimandobs_g(X),
\end{align*}
we arrive at the null hypothesis  
\begin{equation}
\label{eq:nullg}
    \hnull^{\mathcal G}:  \EE_{\prct}\left [ \psi_{\trueg}(Z)\mid \subx \right ]  = 0,\quad \mathrm{for\;some}\; \trueg \in \GG,\quad \prct_{\subx}-\mathrm{almost\;surely}.
\end{equation}
At first glance, testing the null hypothesis in~\Cref{eq:nullg} may seem equivalent to testing equality of conditional means, a problem that has already been extensively studied~\citep{delgado1993testing,neumeyer2003nonparametric,racine2006testing,luedtke2019omnibus,muandet2020kernel}. However, we remark that this equivalence holds only if the function $\trueg$ is known, and to our knowledge, the more realistic scenario where $\trueg$ is unknown has not been previously explored in the literature. 

\subsection{Oracle test statistic}
\label{sec:oracletest}
We now derive a kernelized test statistic for the null hypothesis in~\Cref{eq:nullg}. First, we observe that the hypothesis $\hnull^\GG$ implies an infinite set of unconditional moment constraints, i.e. for any $g \in \GG$, it holds that 
$$\EE_{\prct}\left [ \testrv_g(Z) \mid \subx \right ]  = 0,\;\;\prct_{\subx}-\mathrm{almost\;surely}   \iff \EE_{\prct}\left [ \testrv_g(Z) f(\subx)\right ] = 0,\quad \mathrm{for \;all\;measurable\;}   f.$$
Therefore, 
the validity of testing the RHS would carry over to the validity of testing $\hnull^\GG$. However, testing the RHS of the implication above for all measurable functions is infeasible. Instead,
we can restrict $f$ to be in a reproducing kernel Hilbert space (RKHS). The problem then becomes more tractable since it holds that
\begin{align}
\tstat^2(\psi_g) \defeq \left(\sup _{\|f\|_\FF \leq 1}\EE_{\prct}\left [ \psi_g(Z) f(\subx)\right ] \right )^2 \label{eq:defH} &=  \left \|\EE_{\prct}\left [ \psi_g(Z) k(\subx,\cdot)\right ]  \right \|_\FF^2
\\&= \EE_{\prct}\left [ \psi_g(Z) k(\subx,\tilde{X}^{\J}) \psi_g(\tilde Z) \right ] \nonumber, 
\end{align}
where $k$ is a uniformly bounded reproducing kernel corresponding to a universal RKHS $\mathcal F$~\citep[Definition 4]{steinwart2001influence}, and $\tilde Z$ is an independent copy of $Z$ following the same distribution. In particular, the null hypothesis $\hnull^\GG$ holds if and only if there exists some function $\trueg \in \GG$ such that $\tstat^2(\testrv_{\trueg})=0$, and thus we can construct a valid statistical test based on the oracle statistic $\tstat^2(\psi_{\trueg})$.

\paragraph{A valid test statistic} Given i.i.d. samples $Z_i$ from $\prct$, an unbiased empirical estimate of $\tstat^2(\testrv_g)$ is the cross U-statistic~\citep{kim2024dimension},  defined as  
\begin{equation*}
  \tstathat^2(\testrv_g) \defeq \frac{2}{\nrct} \sum_{i=1}^{\nrct/2} h(Z_i ; \testrv_g), \;\; \mathrm{with}\;\;  h(Z_i; \testrv_g) \defeq \frac{2}{\nrct} \sum_{j=\nrct/2 +1}^{\nrct} \testrv_g(Z_i) k(\subx_i, \subx_j) \testrv_g(Z_j), \;\text{for all}\;g \in \GG .
\end{equation*}
Suppose the null hypothesis $\hnull^\GG$ holds, and fix one function $\trueg \in \GG$ such that
\[
\EE_{\prct}\left [ \psi_{\trueg}(Z)\mid \subx \right ] = 0,\quad \prct_{\subx}\text{-almost surely}.
\]
 The main advantage of the cross U-statistic is that, for $g=\trueg$, it is asymptotically normal under the null hypothesis $\hnull^\GG$ and weak regularity assumptions~(see Theorem~\ref{thm:main}), i.e. as $\nrct \to \infty$ it holds that
 \begin{equation*}
     \sqrt{\frac{\nrct}{2}} \;\frac{\tstathat^2(\testrv_{\trueg})}{\hat \sigma\left(\tstathat^2(\testrv_{\trueg})\right)} \to \gauss \left(0,1\right), 
 \end{equation*}
where $\hat \sigma^2\!\left(\tstathat^2(\testrv_{g})\right)$ is defined as 
\begin{equation*}
\hat \sigma^2\!\left(\tstathat^2(\testrv_g)\right)
\defeq
\frac{2}{\nrct}\sum_{i=1}^{\nrct/2} h(Z_i;\testrv_g)^2
-
\left(\tstathat^2(\testrv_g)\right)^2,
\; \text{for all } g \in \GG.
\end{equation*}
The asymptotic normality result above is stated for an oracle choice $\trueg$, which is infeasible in practice since such a function is unknown. To obtain a computable test statistic, we exploit the fact that under the null hypothesis at least one such function belongs to $\GG$. Therefore, minimizing over $g \in \GG$ can only decrease the absolute value of the standardized statistic, i.e.
\begin{equation*}
\tstatopt \defeq \underset{\g \in \GG}{\min}\left\vert \sqrt{\frac{\nrct}{2}} \frac{\tstathat^2(\psi_g)}{\hat \sigma\left(\tstathat^2(\psi_g)\right)}\right\vert \leq \left\vert  \sqrt{\frac{\nrct}{2}} \frac{\tstathat^2(\psi_{\trueg})}{\hat \sigma\left(\tstathat^2(\psi_{\trueg})\right)}\right\vert.
\end{equation*}
Since the oracle statistic on the right converges in distribution to the absolute value of a standard normal random variable, we can compare $\tstatopt$ to the corresponding half-normal quantiles and obtain an asymptotically valid, albeit potentially conservative, test.

\paragraph{Why not a classic U-statistic?}
We remark that it is not clear how to test the null $\hnull^\GG$ using the classic U-statistic~\citep{serfling2009approximation}, as done in previous works (see e.g.~\citep{hussain2023falsification,demirel2024benchmarking}). The main challenge is that under the null hypothesis $\tstat^2(\psi_{\trueg})=0$,  the U-statistic converges in distribution to a weighted $\chi^2$-statistic. However, estimating the quantiles (needed for a valid test) of this asymptotic distribution via bootstrapping requires knowing the function $\trueg$~\citep{huskova1993consistency}. 
In contrast, our test statistic $\tstatopt$ is bounded by a valid asymptotic pivot, i.e. a function of the data and the unknown function $\trueg$ whose asymptotic distribution does not depend on $\trueg$. Hence, comparing $\tstatopt$ with the quantiles of the half-normal distribution yields an asymptotically valid test.

\subsection{Theoretical guarantees}
\label{subsec:guarantees}
Since, in practice, we do not have access to the signal function $\testrv_g$, we define the finite-sample analogous  as
\begin{equation*}
\hat \testrv_{g}(Z) =  Y \left(\frac{T}{\pi}-\frac{1-T}{1-\pi}\right) - \hatestimandobs_g(X), \quad\text{where}\;\; \hatestimandobs_g(X) \defeq  g(\subx )\hatubobs(X) + \left(1-g\left(\subx\right)\right)\hatlbobs(X),
\end{equation*}
and $\hat\tau_{\pm}^\obs$ is a consistent estimate of $\tau_{\pm}^\obs$ that uses only the observational data $\dataobs$. We can then define our finite-sample test statistic as
\begin{equation}
\label{eq:test}
	\tstathatopt\defeq \underset{\g \in \GG}{\min}\left\vert \sqrt{\frac{\nrct}{2}} \frac{\tstathat^2(\hat \psi_g)}{\hat \sigma\left(\tstathat^2(\hat \psi_g)\right)}\right\vert, 
	~~\text{and the testing function}~~  \hat \phi(\alpha) := \indi \left\{ \tstathatopt\geq z_{1-\alpha}\right\},\end{equation}where $z_\alpha$ is the $\alpha$-quantile of the half-normal distribution.  Below, we provide sufficient conditions for $\test$ to be an asymptotically valid test.  
\begin{thm}[Validity of the test]
\label{thm:main} 
Suppose the null hypothesis $\hnull^\GG$ holds, and fix one function $\trueg \in \GG$ for which the conditional moment restriction in Equation~\eqref{eq:nullg} holds. We make the following assumptions:
\begin{enumerate}
\item[(i)] The variance term is non-zero, i.e.
$    \mathbb E_{\prct} \left [ \testrv^2_{\trueg}(Z)~k^2(\V, \tilde{X}^\J )  ~\testrv^2_{\trueg}(\tilde Z) \right ]>0
$.
\item[(ii)]
The estimates $\hatestimandobs_{\pm}$  satisfy $  \| \estimandobs_{\pm} -  \hatestimandobs_{\pm}\|_{L^2(\mathbb \prct)}  = O_{\pobs}\left(\frac{1}{\sqrt{\nobs}}\right)$, and it holds that 
$  \underset{\nrct,\nobs \to \infty}{\lim} \nrct/\nobs=  0.
$
\end{enumerate}\vspace{-1.5mm}
Then, we have that $$\sqrt{\frac{\nrct}{2}} \frac{\tstathat^2(\hat \psi_{\trueg})}{\hat \sigma\left(\tstathat^2(\hat \psi_{\trueg})\right)}\to \gauss(0,1), \;\;\text{as}\;\; \nrct \to \infty \;\;\text{and}\;\; \nobs \to \infty.$$ 
Hence, 
$\test( \alpha)$ is  a valid asymptotic test at level $\alpha$  for the null hypothesis  $\hnull^\GG$ from Equation~\eqref{eq:nullg}.
\end{thm} 
We refer the reader to~\Cref{apx:proofthm} for a complete proof.

\paragraph{Discussion of assumptions} Assumption~(\textit{i}) is mild and applies to very general settings, e.g. it is satisfied when $Y$ is a non-deterministic random variable. Assumption (\textit{ii}) is stronger and generally only expected to hold when $\nobs\gg\nrct$ and the support of the randomized control trial is contained in the support of the observational study, i.e.  
$\supp(\prct_X) \subseteq \supp(\pobs_X).$
These two conditions are realistic in our setting, as they align with the standard design of observational studies~\citep{franklin2019evaluating, schurman2019framework, he2020clinical}. Further, we remark that previous works either assume oracle access to the functions $\estimandobs_{\pm}$~\citep{hussain2023falsification,demirel2024benchmarking} or impose similar assumptions on the rates~\citep{de2023hidden}.

\paragraph{Power of the test}
While Theorem~\ref{thm:main} only shows asymptotic validity, we further present guarantees for the asymptotic power of the test in~\Cref{sec:power}. 
In particular, in Theorem~\ref{thm:power}, we show that under the alternative hypothesis 
\begin{equation*}
    H_A^{\GG }: \inf_{g \in \GG} \sup _{\|f\|_\FF \leq 1}\EE_{\prct} \left [ \psi_{g}(Z) f(\V)\right ]  >0,
\end{equation*}
the test statistic $\tstatopt$ grows at the typical rate of order $\sqrt{\nrct}$ for a fixed function class $\GG$.  Thus, it yields the same asymptotic power as the existing conditional moment tests~(see e.g. \citep{muandet2020kernel,hussain2023falsification}).


\subsection{A strategy for benchmarking the observational study} Given the theoretical results in this section, we can now introduce our strategy to benchmark observational studies. To do so, we first leverage both tolerance and granularity to estimate an asymptotically valid lower bound on the maximum bias for any subgroup in the observational study, that is $\proxybias \defeq \| \estimandobs -\estimandrct \|_{L^\infty(\prct)}$.

More concretely, we choose as tolerance functions $\estimandobs_{\pm}(X) = \estimandobs(X) \pm \delta$, for some constant $\delta \in \RR^+$, and  we define a data-dependent lower bound on the bias as\begin{align}
\label{eq:deltalb}
\deltalb \defeq \inf_{\delta}\{ \delta : \test(\alpha) = 0\},
\end{align}
where $\test$ depends implicitly on $\delta$ via the tolerance functions and we fix $\J= \{1,\ldots,d\}$. 
Then, under the assumptions in~\Cref{thm:main}, it holds that
$$
\mathbb P \left ( \proxybias \geq \deltalb \right) \geq 1 - \alpha + o(1).
$$
Crucially, to benchmark the observational study, we propose to compare the lower bound on the bias against a critical value, e.g. the minimum bias strength that would explain away the estimated treatment effect in a subgroup of interest. If the lower bound is greater than the critical value, we discard the conclusions drawn from the observational study. In~\Cref{sec:rwexp}, we will demonstrate that our strategy yields conclusions consistent with current epidemiological knowledge using real-world data from the Women's Health Initiative.

\paragraph{Limitations} 
We remark here that the lower bound defined in~\Cref{eq:deltalb} is optimistic, as there are two potential sources of looseness. First, a lack of power in our testing procedure can result in a lower bound far from $\proxybias$. However, at least in principle, this gap can be improved by future work on more powerful tests. Second, the bias could be arbitrarily high outside the support of the randomized trial, that is $\| \estimandobs -\cateobs \|_{L^\infty(\pobs)} > \|\estimandobs -\cateobs \|_{L^\infty(\prct)} =   \truebias$. Unfortunately, we cannot reduce this gap without making further assumptions on the bias structure that would allow us to extrapolate beyond the randomized trial support.

\section{Semi-synthetic experiments}
\label{sec:exp}

In this section, we evaluate our test and the resulting bias lower bound in finite-sample semi-synthetic experiments.

\subsection{Experimental setting}

\paragraph{Dataset} We evaluate our testing procedure on a semi-synthetic dataset derived from a real-world randomized trial: Hillstrom's MineThatData Email dataset \citep{hillstrom2008}. The Hillstrom dataset contains records of 64,000 customers who made purchases online within the last twelve months. We consider a combined treatment group, which constitutes approximately 66\% of the dataset, and a control group. The outcome represents the dollars spent in the two weeks post-campaign. The dataset provides information on individual annual spending, newcomer status, and geographical location, among others. We normalized continuous features and one-hot-encoded categorical features, resulting in a 13-dimensional dataset. By default, we use 80\% of the full dataset as the observational study ($\obs$) and the remaining 20\% as the randomized trial ($\rct$).

\paragraph{Bias model} We consider three different models for the bias between studies, given by $\truebias(x)=\cateobs(x)-\estimandobs(x)$, for all $x \in \XX$. In Scenario 1, we consider a single subgroup with a constant bias of $\truebias=60$, while the rest of $\obs$ remains unbiased. In Scenario 2 (\Cref{fig:heatmap_scenario2}), we add biases of varying magnitudes across 12 subgroups defined by combinations of the binary features \texttt{newbie} and \texttt{mens} and the categorical feature \texttt{channel}. The largest bias is $\truebias=60$, and it affects only 12\% of the observational dataset.
The subgroup biases roughly cancel each other out on average, resulting in an average bias close to zero, i.e. $\EE_{\pobs}[ \truebias(X)] \approx 0$. Finally, in Scenario 3 (\Cref{fig:heatmap_scenario3}), we model the bias as a quadratic polynomial of the feature \texttt{history}, sampling different coefficients for the two values of \texttt{newbie} from a normal distribution.

\paragraph{User-defined tolerance and baselines} 
We refer to the testing function proposed in this paper as $\catetest$, and we instantiate it using constant upper and lower bounds for the tolerance function, as described in Example 1 from~\Cref{sec:hte}~($\estimandobs_{\pm}(X) = \estimandobs(X) \pm \delta$ for some constant $\delta \in \RR^+$). We compare our test against $\atetest$, a t-test for the null hypothesis that average treatment effects between the studies differ at most $\delta$ (this is equivalent to the test $\catetest$ when the selected subset of features $\mathcal J$ is the empty set). For both tests, we can compute the lower bound on the bias $\deltalb$, as defined in~\Cref{eq:deltalb}. Note that while our method allows us to select a subset of features $\subx$ that are interesting for the treatment effect heterogeneity, we use the full feature set in all our semi-synthetic experiments. We thus show the effectiveness of our test even when considering a relatively large set of features, and we expect power to improve when considering a smaller subset; see, e.g. the ablation studies for $\subx$ in \Cref{sec:ablation_subset}.

\paragraph{Implementation} We use the Laplacian kernel with a scale of 1.0 to compute our test statistic $\catetest$. We perform gradient descent for 6000 epochs using the \texttt{Adam} optimizer from the JAX-based library \texttt{optax} with its default hyperparameters and record the smallest test statistic. As function class $\GG$, we consider linear functions and two multilayer perceptrons (MLPs), one \textit{small} and one \textit{large}, with hidden layer widths of 10 and 100-50-10-5 neurons, respectively. For the linear function and the small MLP, we set the learning rate to 0.1, and for the large MLP, we set it to 0.01. For the test  $\atetest$, we use 500 bootstrap samples to estimate the variance of the test statistic.

\subsection{Experimental results}
\begin{figure*}
\centering
    \begin{subfigure}[b]{0.245\textwidth}
        \centering
        \includegraphics[width=\textwidth]{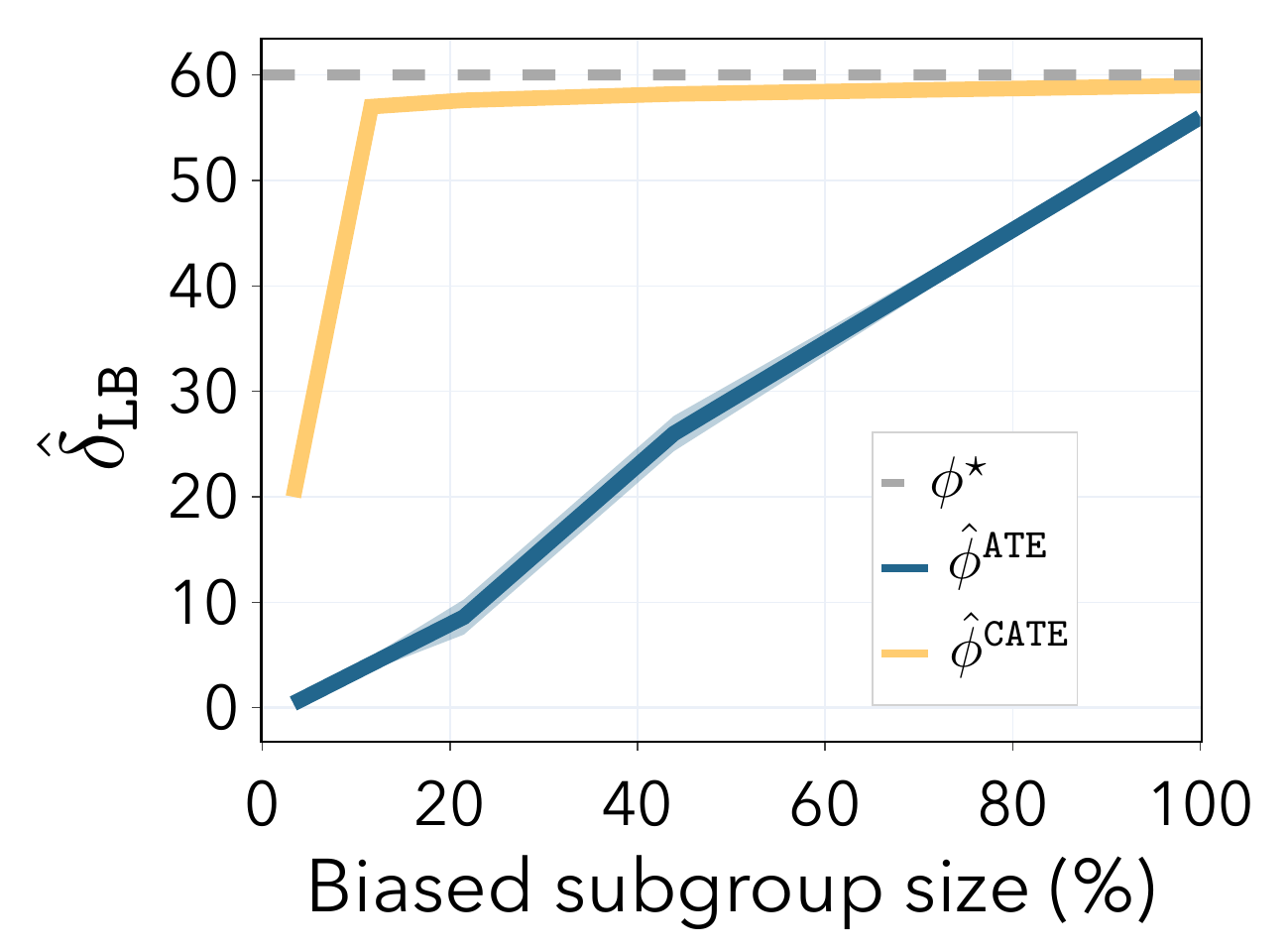}
        \caption{Scenario 1
        }
      \label{fig:abl_bias_group}
    \end{subfigure}
    \hfill
    \begin{subfigure}[b]{0.245\textwidth}
        \centering
        \includegraphics[width=\textwidth]{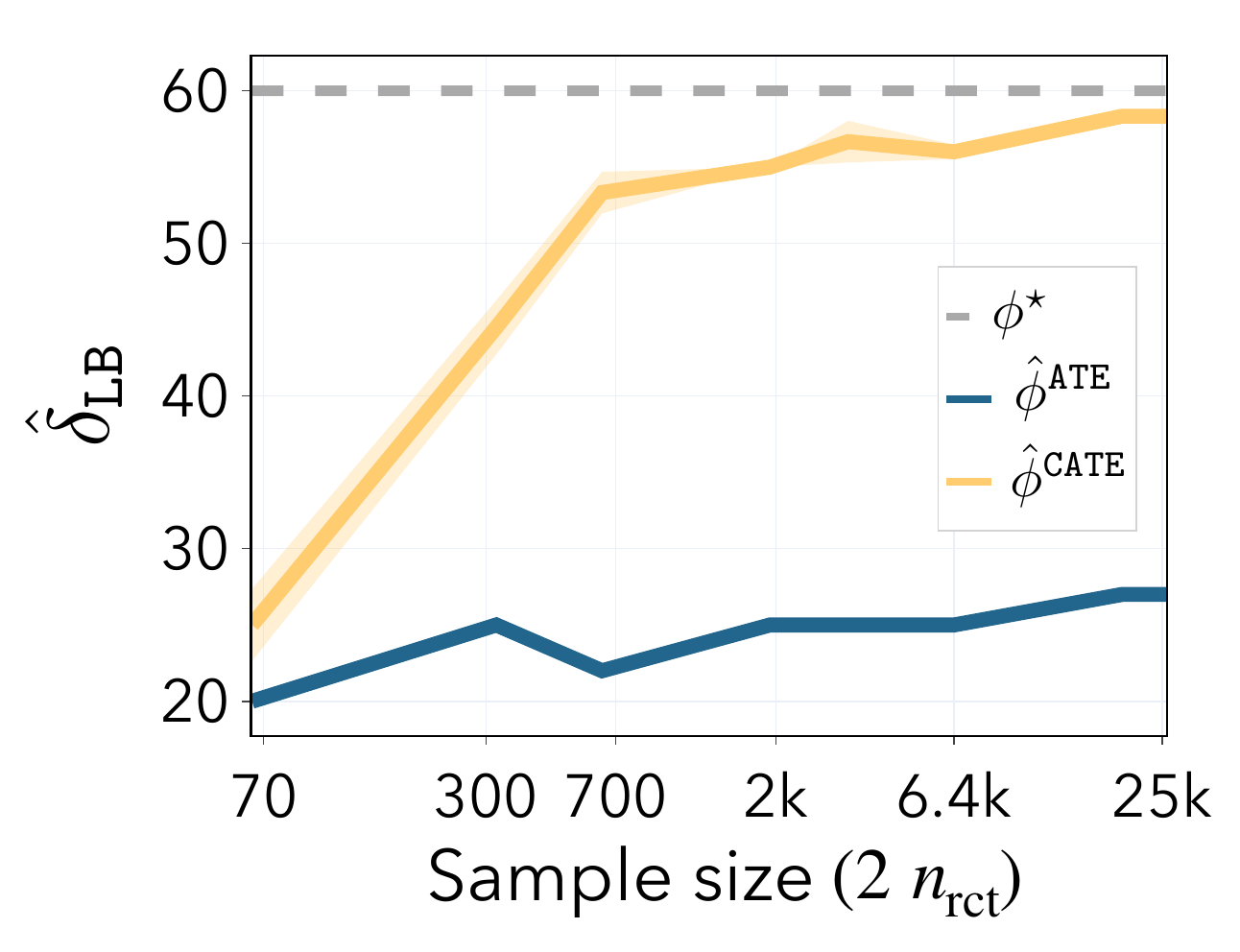}
        \caption{Scenario 1}
        \label{fig:abl_rct_size}
    \end{subfigure}
    \hfill
    \begin{subfigure}[b]{0.245\textwidth}
        \centering
        \includegraphics[width=\textwidth]{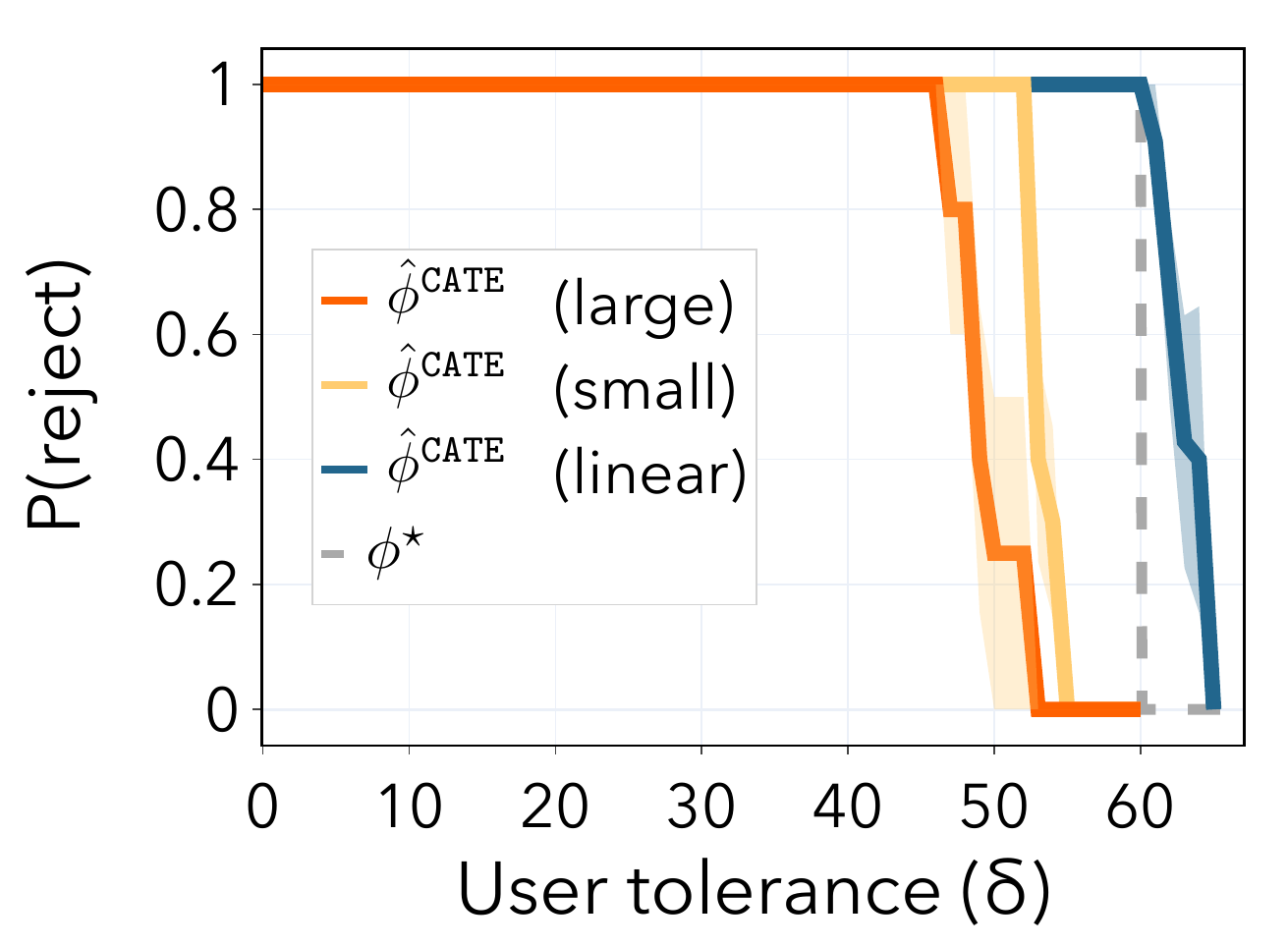}
        \caption{Scenario 2}
        \label{fig:abl_function_class}
    \end{subfigure}
    \hfill
    \begin{subfigure}[b]{0.245\textwidth}
        \centering
        \includegraphics[width=\textwidth]{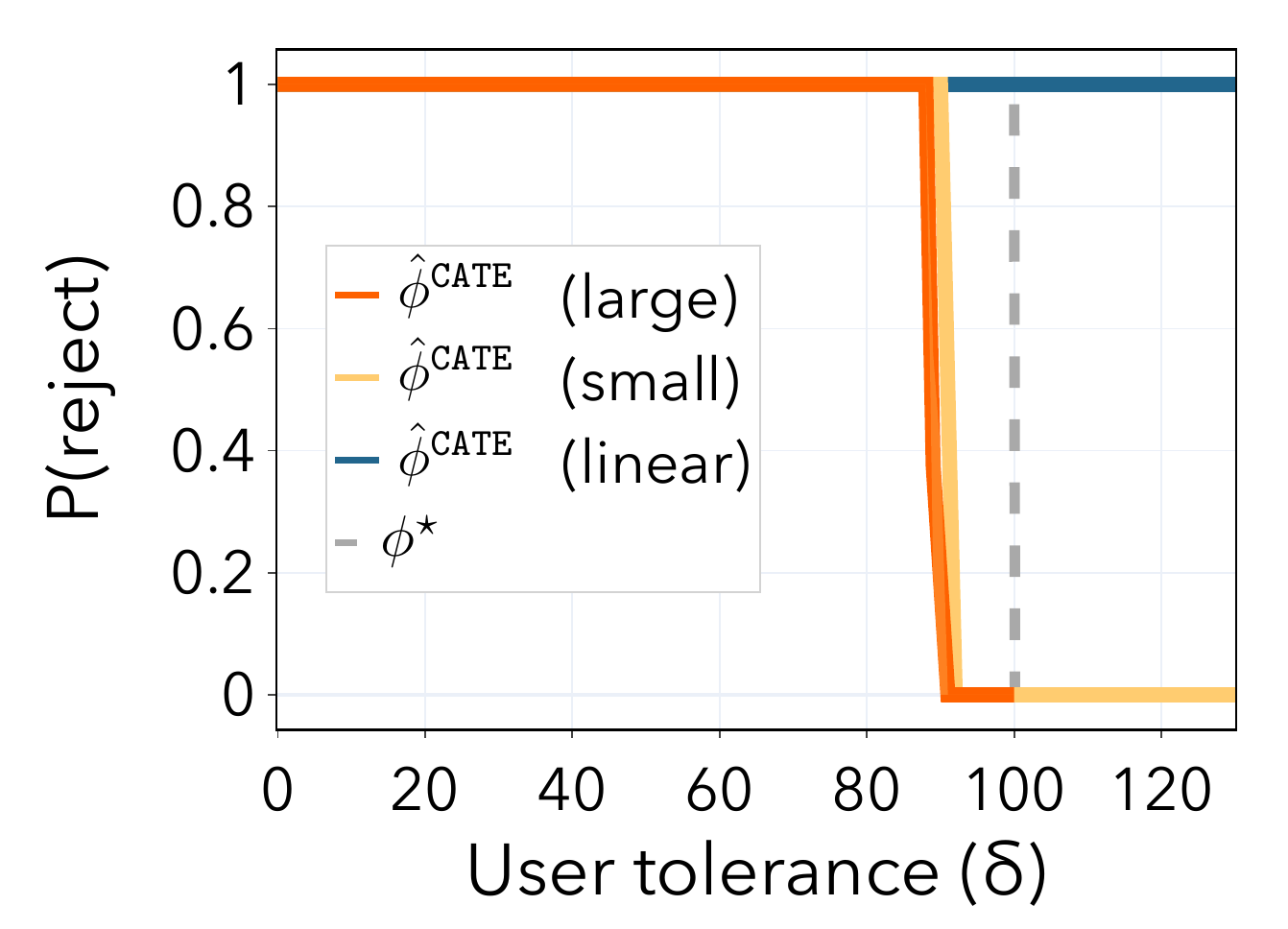}
        \caption{Scenario 3}
        \label{fig:abl_function_class_hard}
    \end{subfigure}\caption{
For all the plots: the significance level is set at $\alpha=0.05$, $\phi^\star$ denotes the oracle test, which rejects for $\delta<\truedelta$. (a-b) Scenario 1, comprising a single subgroup with a constant bias $\truedelta=60$: we plot the bias lower bound $\deltalb$ as a function of (a) the biased subgroup percentage w.r.t. total sample size and (b) the randomized trial sample size. (c-d) Probability of rejection for different function classes $\GG$ as a function of the user-specified tolerance $\delta$ for (c) Scenario 2 (\Cref{fig:heatmap_scenario2}) based on 12 subgroups with different biases and (d) Scenario 3 (\Cref{fig:heatmap_scenario3}) based on a quadratic polynomial bias. We report mean and standard error over 5 runs. The coefficients for the polynomial bias are fixed across runs. 
}
    \label{fig:hillstrom_exp}
\end{figure*}

We now discuss our experimental results, depicted in \Cref{fig:hillstrom_exp}. We first conduct ablation studies for Scenario 1, with only one subgroup having a constant bias of $\delta^*=60$. We study the effect of the biased subgroup size (\Cref{fig:abl_bias_group}) and the randomized trial sample size (\Cref{fig:abl_rct_size}) on the lower bounds $\deltalb$ obtained from our test $\catetest$ and the baseline $\atetest$. Next, we assess the validity and power of our test $\catetest$ in two more complex settings: Scenario 2 (\Cref{fig:abl_function_class}) and Scenario 3 (\Cref{fig:abl_function_class_hard}).  An important consideration is the selection of the function class $\GG$ in practice; it should be sufficiently large to contain $\trueg$, but overly large function classes may result in a more complex optimization problem. Thus, we also conduct ablation studies for $\GG$. Our results show that granularity significantly improves the power of the test and, consequently, the estimated lower bound on the bias: $\catetest$ consistently outperforms the baseline across all scenarios and demonstrates robustness w.r.t. the choice of function class in the ablation studies. 

\paragraph{Effect of biased subgroup and rct sample sizes} \Cref{fig:abl_bias_group} shows that our test yields an average lower bound $\deltalb$ smaller and close to the true maximum bias $\truedelta$. This implies that the test remains valid and exhibits significant power, even when the biased subgroup represents roughly 14\% of the observational dataset. In contrast, $\atetest$  experiences a significant drop in power as the proportion of biased data points decreases. Such behavior is expected since $\atetest$ only tests for the difference of averages, and it cannot detect bias in small subgroups, i.e. it is not granular. In \Cref{fig:abl_rct_size}, we add a constant bias of 60 to 44\% of the observational data points and study the effect of the randomized trial sample size. While our test suffers more than $\atetest$ from a decrease in the sample size due to the use of kernels, it always yields higher power, even in the very small sample size regime with 70 data points. These results show the importance of granularity: even in simple settings, $\atetest$  can fail to flag significantly biased datasets, in contrast to our method.

\paragraph{Validity and power in complex scenarios} 
\Cref{fig:abl_function_class} and \Cref{fig:abl_function_class_hard} show the validity and power of our testing procedure for Scenario 2 (illustrated in~\Cref{fig:heatmap_scenario2}) and Scenario 3 (illustrated in~\Cref{fig:heatmap_scenario3}), respectively. In both scenarios, if we use a neural network to approximate the bias function, our test remains valid and shows very high power since it rejects the null hypothesis at values of $\delta$ close to the true bias $\truedelta$.  
\paragraph{Effect of misspecified function class}  Notably, when $g$ is modeled with a linear function, our test loses its validity, rejecting values of $\delta$ that are larger than the true bias. Such behavior is expected as the chosen function class $\GG$ lacks the complexity necessary to capture the true bias model. Nevertheless, we observe that the \textit{small} network with one hidden layer is already sufficient. Further, significantly increasing the complexity -- the \textit{large} network has approximately 45 times more parameters than the \textit{small} one -- still yields high power. Therefore, we recommend practitioners to be conservative in their choice of function class to ensure validity, even if it might come at the potential cost of some power and a more complex optimization problem. Moreover, although we cannot guarantee convergence to a global optimum, given the non-convexity of the problem for complex function classes, we show that the optimization procedure is stable and consistently reaches the same solution in~\Cref{apx:opt}.

\section{Real-world experiments}
\label{sec:rwexp}

In this section, we provide a concrete application of the benchmarking framework using the Women's Health Initiative (WHI) study. We show the strengths of our testing procedure and how tolerance and granularity are necessary for effective benchmarking.

\subsection{The WHI controversy}
The WHI study included a randomized trial and an observational study that investigated the use of hormone therapy~(HT) for preventing common sources of mortality among postmenopausal women, including cardiovascular disease, cancer, and fractures~\citep{anderson2003implementation}.

\paragraph{To HT, or not to HT} The initial results of the WHI study in 2002 led to fear and confusion regarding the use of hormone therapy (HT) after menopause, resulting in a dramatic reduction in prescriptions for HT  around the world. Although in 2002, it was stated that HT increases the risk of coronary heart disease (CHD) for all women, subsequent studies clearly showed that younger women close to menopause can benefit from HT. 
 Indeed, for at least 2 decades before the WHI study, observational studies had suggested that HT reduces the risk of CHD~\citep{stampfer1991estrogen,henderson1991decreased,grady1992hormone,grodstein2000prospective}. Further, subsequent randomized trials have continued demonstrating the benefits of HT when started early in young women close to menopause~\citep{hodis2016vascular,taylor2017effects}. To date, the consensus among epidemiologists is that hormone therapy reduces the risk of CHD in women aged less than 60 years and within 10 years of menopause; see e.g. the current guidelines for menopausal hormone therapy~\citep{lee20202020}.

\paragraph{Limitations of the WHI randomized trial} The main issue with the randomized trial from the WHI study is that younger women's cardiac events are relatively rare. Indeed, not only would it have been prohibitively expensive to conduct a randomized trial exclusively in younger women, but it would have also taken many years to accumulate enough events to reach statistical significance. Hence, the trial lacked enough events to reach statistical significance on the subgroup of interest. On the other hand, the average treatment effect (over all the patients in the trial) suggested an increase in CHD risk because the majority of cardiac events came from older women, and epidemiologists concluded that HT is harmful to all women.  
 
  \paragraph{Benchmarking can help!} It has been argued that in the 10 years since the WHI study, many women have been denied HT, significantly disadvantaging a generation of women~\citep{sturdee2011updated}. The natural question is, thus, if, going back in time,  benchmarking the observational study could have prevented such a turn of events.  
  Indeed, this is the perfect setting to test our methodology, as we would like to ask the question:
 \begin{align*}\quad & \emph{Is the bias in the observational study enough to explain away } \\&\emph{\;\;\;the benefits of HT in young women close to menopause?}
 \end{align*}
 In what follows, we show that answering such a question requires a statistical test that offers tolerance.  
 Further, even though we cannot demonstrate that granularity is necessary in this concrete example\footnote{To do so, we would need to know a small biased subgroup in the observational study and show that only the tests with granularity detect the bias. Unfortunately, we are unaware of subgroups that were found to be biased in the WHI study.}, we stress that it is equally important in practice. This is especially true with respect to age and time since the start of menopause, as the tests without granularity can fail to detect subgroup bias that cancels on average, as shown in our semi-synthetic experiments.

\begin{table}
\centering

\caption{The significance level is set at $\alpha=0.05$. $ \deltact$ is the amount of bias that would explain away the positive effect of HT in young women close to menopause. $ \deltalb$ is the maximum bias detected in the observational study. $\atetestzero$ and $\catetestzero$ denote the test from~\Cref{eq:test} with tolerance functions $\estimandobs_{\pm}(x) = \estimandobs(x)$ for all $x \in \XX$, when $\mathcal J$ is the empty set or the entire set of features, respectively.}
\label{table:whi}
\begin{tabular}{cccccc}
\toprule
\multirow{1}{*}{Statistical tests} & \multicolumn{1}{c}{$\catetest$} & \multicolumn{1}{c}{$\atetest$}&  \multicolumn{1}{c}{$\catetestzero$}&   \multicolumn{1}{c}{$\atetestzero$} \\ 
\midrule
$\deltact$ & $0.32$  & $0.32$ &  $0.32$ & $0.32$ \\
$\deltalb$  & $\mathbf{0.25}$  &    $0.11$  &\xmark & \xmark\\ 
\midrule
Reject the study  & $\color{mygreen} \mathbf 0$   & $\color{mygreen} \mathbf 0$ & $\color{pierLink} \mathbf 1 $ &  $\color{pierLink} \mathbf 1 $ \\
\bottomrule
\end{tabular}
\end{table}

\subsection{Experimental results}
Linking back to our question of interest, we demonstrate how our method can provide a correct answer, i.e. one that aligns with the epidemiology literature. A natural way to do so is to first estimate from the available data the amount of bias that would explain away the treatment effect on the group of interest, defined as 
$$
\deltact \defeq \Big|\EE_{\pobs} \left [  \estimandobs(X) \mid X \in G \right ] \Big| .
$$
In essence, the critical value quantifies the minimum strength of bias for which positive and negative values of treatment effect are reasonable, thereby invalidating the observational study results\footnote{Note that other choices for the critical value are possible, and practitioners should determine the most appropriate one given the specific context.}. In our example, the group $G$ is defined as young women (age $\leq 60$) who are close to menopause ($\leq 10$ years).

Similarly to the semi-synthetic experiments, we instantiate the tolerance functions using constant upper and lower bounds, i.e. $\estimandobs_{\pm}(X) = \estimandobs(X) \pm \delta$ for some constant $\delta \in \RR^+$. We compute the lower bound $\deltalb$ on the maximum amount of treatment effect bias in the observational study, as defined in~\Cref{eq:deltalb}.
We remark that this quantity can be computed only for tests that allow some tolerance. Then, our decision-making procedure will flag the observational study as invalid if $\deltalb \geq \deltact$.

\paragraph{Experimental details} We consider a binary-valued outcome: the presence of coronary heart disease
within the follow-up period.
 We choose as covariates $X$ the basic adjustment variables used in many existing analyses, and we further limit patients to those who were not current users of HT at the time of enrolment, as the duration of HT use has been found to have a substantial impact on treatment effects~\citep{prentice2005combined,vandenbroucke2009hrt}. We refer to~\Cref{apx:whi_exp} for complete experimental details. 

We now present evidence that our procedure yields conclusions that align with the epidemiology literature. It avoids issuing false alarms when the bias is negligible (tolerance) and detects a larger amount of bias, as it is more powerful than tests based on average treatment effect (granularity).

\paragraph{Results} In~\Cref{table:whi}, we show the result for all the statistical tests on the WHI study. First, we observe that both tests that allow for tolerance correctly do not flag the study, while $\catetestzero$ and $\atetestzero$ do. This difference shows the importance of tolerance for distinguishing between small and large amounts of bias. Second, we observe that the lower bound on the bias is larger for the test with granularity $\catetest$. Such behavior is expected and shows the importance of granularity to detect bias that would otherwise go unnoticed using the test without any granularity $\atetest$.

\section{Related work}

Given the challenges associated with estimating treatment effects using non-randomized data, several works propose to detect bias in the treatment effect estimated from observational data by leveraging randomized trials~\citep{viele2014use,yangelastic, morucci2023double, gao2023pretest}, multiple observational studies~\citep{karlsson2023detecting}, or negative controls~\citep{lipsitch2010negative,donald2014testing,de2014testing,sofer2016negative}. In particular, when randomized data is available, they introduce statistical tests for the null hypothesis 
\begin{align}
\label{eq:atetest}
\hnullate: \EE_{\pxrct}\left [\estimandrct(X)\right] = \EE_{\pxrct}\left[\estimandobs (X)\right].
\end{align}
Rejecting $\hnull$ implies that either the treatment effect estimate from the observational study is biased or the transportability assumption is violated, i.e.  $\caterct(x) \neq \cateobs(x)$ for some $x \in \XX$. However, the null hypothesis in~\Cref{eq:atetest}  does not allow tolerance or granularity. Thus, it suffers from two major limitations: it rejects observational studies with negligible treatment effect bias 
and it cannot detect bias in small subgroups or individuals. 
   In the following, we present existing statistical tests designed to offer either tolerance or granularity, and we describe how our method generalizes them.


 \paragraph{Statistical tests with tolerance} 
One way to address the restrictiveness of previous statistical tests and reduce false rejections is to incorporate some tolerance. More formally,  given some user-specified tolerance functions $\estimandobs_{\pm}$, \citet{yangelastic,de2023hidden}
 propose a test for the null hypothesis
\begin{align}
\label{eq:atetol}
\hnull: \EE_{ \pxrct}\left[\estimandrct(X)\right] \in \left[ \EE_{ \pxrct}\left[\lbobs(X)\right], \EE_{\pxrct}\left[\ubobs(X)\right]\right],
\end{align}
where $\lbobs(x) \leq \estimandobs(x) \leq \ubobs(x)$ for all $x \in \XX$. For instance, if we choose sensitivity analysis bounds as tolerance functions and assume transportability, we can test for the presence of unobserved confounding above a certain strength.  However, current statistical tests with tolerance are not granular: large biases in small subgroups can remain undetected. In contrast, our null hypothesis in~\Cref{eq:catetolnull} allows granularity and recovers existing tests with tolerance when the subset of features $\mathcal J = \emptyset$.


\paragraph{Statistical tests with granularity}
Several works have addressed the lack of granularity. \citet{hussain2022falsification} compare group-level treatment effects using pre-specified subgroups; however, this approach suffers from multiple testing issues. More recently, \citet{hussain2023falsification} propose a kernel test for the null hypothesis \begin{align}
\label{eq:catetest}
    \hnull:  \estimandrct(X) = \estimandobs (X), \quad  \pxrct-\mathrm{almost\;surely.}
\end{align}
 The main advantage of such a test is that it can detect bias in arbitrarily fine-grained subpopulations without suffering from multiple testing corrections. Further, \citet{demirel2024benchmarking} extend it to account for right-censored outcomes. However, all the statistical tests with granularity fall short of incorporating tolerance functions. In contrast, our null hypothesis in~\Cref{eq:catetolnull} allows tolerance 
and recovers $\hnull$ in~\Cref{eq:catetest} when the tolerance functions are the same ($\lbobs(x) = \ubobs(x)$ for all $x \in \XX$), and we set $\J = \{1,\ldots,d\}$.

\paragraph{Combining data for  estimation}
 In the presence of observational and randomized data, an alternative to testing involves estimating the bias and correcting for it, ultimately leading to a more accurate treatment effect estimate
~\citep{kallus2018removing,yang2020improved,wu2022integrative,yangelastic,rosenman2023combining,yuwen2023enhancing}. These approaches are promising when the support of the two studies is the same, as they can reduce the variance of the treatment effect estimates by pooling the data.
The work of \citet{cheng2021adaptive} is particularly related to ours, where the authors use kernel regression to estimate the treatment effect conditional on a subset of features. 
However, a caveat of bias correction is that it requires matching support of both studies: when the supports are different, learning the bias requires strong parametric assumptions for extrapolation. In contrast, statistical tests aim to identify flawed observational studies; see, e.g. \cite{forbes2020benchmarking}. This task is feasible even in settings where the supports do not match, as it is enough to detect differences in the common support of the two studies.  

\section{Limitations and future work}
Our approach shares limitations with other methods that rely on kernels for testing. Most notably, the curse of dimensionality can be a significant problem given the small sample size of randomized trials. In addition, the benchmarking strategy is optimistic; outside the common support of the two studies, the bias could be arbitrarily higher than our lower bound $\deltalb$. 

Our discussion suggests several important directions for future research. 
For example, our test could be adapted to the scenario where multiple observational datasets may be available but no randomized trials. Further, in settings where  $\nrct \approx \nobs$, or the tolerance functions $\estimandobs_\pm$ are difficult to learn, Assumption (\textit{ii}) in Theorem~\ref{thm:main} may be unrealistic. One way to overcome this limitation is to construct a semiparametric efficient test statistic that combines multiple nuisance functions to relax the required assumptions on the approximation quality of the individual nuisance functions. Finally, applying and validating our method in more real-world scenarios presents an exciting avenue for future work.

\subsection*{Acknowledgements}
PDB was supported by 
the Hasler Foundation grant number 21050. JA was supported by the ETH AI Center. KD was supported by the ETH AI Center and the ETH Foundations of Data Science.


\bibliography{main}
\bibliographystyle{plainnat}
\clearpage
\section*{Appendices}
The following appendices provide deferred proofs, experiment details, and ablation studies.
\appendix
\DoToC
\clearpage
\hypersetup{
    colorlinks,
   linkcolor={pierLink},
    citecolor={pierCite},
    urlcolor={pierCite}
}
\section{Methodology}
For the sake of clarity, we write $n \defeq \nrct/2$ throughout this section. Moreover, we let $\mathbb P$ denote the joint law of $(\datarct,\dataobs)$, and take $\EE\left[\cdot\right]$ with respect to $\mathbb P$ unless otherwise specified.

\subsection{Proof of Theorem~\ref{thm:main}}
\label{apx:proofthm}
Suppose the null $\hnull^\GG$ holds, and fix one function $\trueg \in \GG$ such that
\[
\EE_{\prct}\left[\psi_{\trueg}(Z)\mid \subx\right] = 0,\quad \prct_{\subx}-\text{almost surely}.
\]
 We begin  with the simple observation that
\begin{equation*}
\underset{\g \in \GG}{\min}\left\vert  \obj{\hat \testrv_{g}} \right\vert \leq \left\vert  \obj{\hat \testrv_{\trueg}} \right\vert,
\end{equation*}
which holds under the assumption that $\trueg \in \mathcal G$. Thus, asymptotic validity of $\test$ follows when showing that the RHS converges in distribution to an absolute normal distribution.

We first establish the oracle result. Under Assumption~(\textit{i}) in~\Cref{thm:main}, Theorem~4.2 from~\cite{kim2024dimension} gives the asymptotic normality:
\begin{align*}
  \obj{ \testrv_{\trueg}} \to \mathcal N(0,1), \;\; \text{as}\;\; n \to \infty. 
\end{align*}
Moreover, as a consequence of Equations~(55)--(57) in the proof of Theorem~4.2 from~\cite{kim2024dimension}, we have that
\begin{equation*}
    \frac{1}{n \varhat{ \psi_{\trueg}}} = O_{\prct}(1).
\end{equation*}
To transfer this result to the estimated statistic, the key ingredient is to show that the plug-in numerator and variance estimator are asymptotically equivalent to their oracle counterparts under $\mathbb P$:
\begin{equation}
\label{eq:proof_main_goal}
\begin{aligned}
\left | n\tstathat^2(\hat \psi_{\trueg})- n\tstathat^2(\psi_{\trueg}) \right|
&= o_{\mathbb P}(1), \\
\left|n\varhat{\hat \psi_{\trueg}}-n\varhat{\psi_{\trueg}} \right|
&= o_{\mathbb P}(1).
\end{aligned}
\end{equation}
Since $\obj{\hat{\testrv}_{\trueg}}$ is obtained from these two quantities through a continuous map, Slutsky's theorem yields
\begin{equation*}
    \obj{ \hat{\testrv}_{\trueg}} \to \gauss(0,1), \;\;  \text{as}\;\;n \to \infty \;\;\text{and}\;\;\nobs \to \infty, 
\end{equation*}
and the statement in Theorem~\ref{thm:main} follows. It now remains to prove Equation~\ref{eq:proof_main_goal}.

\subsubsection*{Proof of statement in Equation~\ref{eq:proof_main_goal}} 
\label{apx:eq12}
We begin by defining the error term 
\begin{align*}
    \triangle := \hat \testrv_{\trueg}(Z)- \testrv_{\trueg}(Z) = -\trueg(\subx)\left(\hatestimandobs_+(X)  - \estimandobs_+(X) \right)   - (1-\trueg(\subx)) \left( \hatestimandobs_-(X) - \estimandobs_-(X) \right),
\end{align*}
and we denote with $\triangle_i$ the corresponding i.i.d. samples (conditional on $\dataobs$). We restate the definition of the mean and variance terms here. Formally, we split the dataset $\datarct$ equally into two folds, $\II_1$ and $\II_2$, of size $n$ and obtain
\begin{align*}
n\tstathat^2(\hat \psi_{\trueg}) &= \frac{1}{\sqrt{n}} \sum_{i \in \II_1} \hat \testrv_{i}  \frac{1}{\sqrt{n}}  \sum_{j \in \II_2}  k(\V_i, \V_j)   \hat \testrv_{j},
\\
n\varhat{\hat \testrv_{\trueg}} &= 
   \frac{1}{n} \sum_{i \in \II_1} \hat \psi_{i}^2 \left(\frac{1}{\sqrt{n}} \sum_{j \in \mathcal I_2} k(\V_i, \V_j) \hat \psi_j\right)^2 - 
  \left(\sqrt{n} \tstathat^2(\hat \psi_{\trueg})\right)^2,
\end{align*} 
where we use the shorthand 
$\hat \psi_i := \hat \psi_{\trueg}(Z_i)$ and $\psi_i := \psi_{\trueg}(Z_i)$. Further, we define the datasets obtained from the two splits as $D_1^{\rct} \defeq (X_i, T_i, Y_i)_{i\in \II_1}$, and  $D_2^{\rct} \defeq (X_j, T_j, Y_j)_{j\in \II_2}$.

\paragraph{Preliminary step: auxiliary quantities and bounds}

Let
$r_n^2 \defeq \EE\left[\triangle^2\mid \dataobs\right].$ By Assumption~$(ii)$ in Theorem~\ref{thm:main}, we have that
\begin{equation*}
    r_n^2  \leq   2 \|\hatestimandobs_+  - \estimandobs_+\|^2_{L_2(\prct)} + 2 \|\hatestimandobs_-  - \estimandobs_-\|^2_{L_2(\prct)} = O_{\pobs}\left(\frac{1}{\nobs}\right), 
\end{equation*}
where the probability $\pobs$ is over the dataset $\dataobs$ used to train $\hatestimandobs_\pm$. Since $n=\nrct/2$ and $\nrct/\nobs\to 0$, it follows that
\begin{equation}
    \label{eq:kernelrates}
r_n=o_{\pobs}(1),
\qquad
\sqrt{n}\,r_n=o_{\pobs}(1),
\qquad
nr_n^2=o_{\pobs}(1).
\end{equation}
We next define, for every $x^\J \in \supp(\pxrctj)$,
\begin{align*}
      \tau_1(x^\J)  &:= \frac{1}{\sqrt{n}}  \sum_{i \in \II_1}  k(\V_i, x^\J)   \triangle_{i},\;\;
      \tau_2(x^\J)  := \frac{1}{\sqrt{n}}  \sum_{j \in \II_2}  k(x^\J, \V_j)   \triangle_{j}.
\end{align*}
We will repeatedly make use of the following bounds:
\[
\EE\left[\tau_2(\V)^2\mid \dataobs\right]=o_{\pobs}(1),
\qquad
\EE\left[\tau_1(\V)^2\mid \dataobs\right]=o_{\pobs}(1).
\]
Indeed, conditioning on $\dataobs$, and using that the pairs $(\V_j,\triangle_j)_{j\in\II_2}$ are conditionally i.i.d.,
\begin{align}
\EE\left[\tau_2(\V)^2\mid \dataobs\right]
&\lesssim
\EE\left[\triangle^2\mid \dataobs\right]
+
n\,\EE\!\left[\EE\left[k(\V,\tilde\V)\tilde\triangle\mid \V,\dataobs\right]^2\middle|\dataobs\right]
 \notag\\
&\lesssim
r_n^2
+
n\,\EE\!\left[\EE\left[\tilde\triangle^2\mid \V,\dataobs\right]\middle|\dataobs\right]\nonumber
\\&=
r_n^2 + n r_n^2
=
o_{\pobs}(1). \label{eq:tau2_bound}
\end{align}
where $(\tilde\V,\tilde\triangle)$ is an independent copy of $(\V,\triangle)$, and we used Cauchy-Schwarz together with boundedness of the kernel and \Cref{eq:kernelrates}. By symmetry, the same argument applies to $\tau_1$.

We will also use two bounds for $$\bar\tau_2(x^\J)  := \frac{1}{\sqrt{n}}  \sum_{j \in \II_2}  k(x^\J, \V_j)   \psi_{j}.$$ Let $(\tilde \V,\tilde{\psi}_{\trueg})$ be an independent copy of $(\V,\psi_{\trueg})$. For every $x^\J \in \supp(\pxrctj)$, using that $(\V_j,\psi_j)_{j\in\II_2}$ are conditionally i.i.d., we have
\begin{equation}
\label{eq:bartau2_conditional}
\begin{aligned}
\EE\left[\bar\tau_2(x^\J)^2\mid x^\J,\dataobs\right]
&=
\EE\left[k(x^\J,\tilde \V)^2\tilde{\psi}_{\trueg}^2\mid x^\J,\dataobs\right]
\\
&\le
\|k\|_\infty^2\|\psi_{\trueg}\|_\infty^2.
\end{aligned}
\end{equation}
where the equality uses the null hypothesis in \Cref{eq:nullg}, since
\[
\EE\left[k(x^\J,\tilde \V)\tilde \psi_{\trueg}\mid x^\J,\dataobs\right]
=
\EE\left[k(x^\J,\tilde \V)\EE\left[\tilde \psi_{\trueg}\mid \tilde \V,\dataobs\right]\mid x^\J,\dataobs\right]
=
0.
\]
Moreover, we also establish that 
\begin{equation}
\label{eq:black_delta_bartau2_bound}
\begin{aligned}
\EE\left[\triangle^2\bar\tau_2(\V)^2\mid \dataobs\right]
&=
\EE\!\left[\triangle^2\,\EE\left[\bar\tau_2(\V)^2\mid \V,\dataobs\right]\middle|\dataobs\right]
\\
&\le
\|k\|_\infty^2\|\psi_{\trueg}\|_\infty^2\,\EE\left[\triangle^2\mid \dataobs\right]
\lesssim
r_n^2 = o_{\pobs}(1),
\end{aligned}
\end{equation}
where we used \Cref{eq:kernelrates,eq:bartau2_conditional}.
We are now ready to show the convergences in Equation~\eqref{eq:proof_main_goal}.

\paragraph{Step 1: controlling $\tstathat^2(\hat \psi_{\trueg})$}
We first control the mean term $\tstathat^2(\hat \psi_{\trueg})$. We show that
\begin{equation}
\label{eq:control_of_mean}
    \left\vert n\tstathat^2(\hat \psi_{\trueg}) - n\tstathat^2( \psi_{\trueg}) \right\vert = o_{\mathbb P}(1).
\end{equation}
We decompose the difference into the following three terms:
\begin{align*}
n\tstathat^2(\hat \psi_{\trueg}) - n\tstathat^2( \psi_{\trueg})
&= \underbrace{ \frac{1}{\sqrt{n}} \sum_{i \in \II_1} \psi_{i} \tau_2(\V_i)}_{=:T_1}
+ \underbrace{ \frac{1}{\sqrt{n}} \sum_{j \in \II_2} \psi_{j} \tau_1(\V_j) }_{=: T_2} + \underbrace{
\frac{1}{\sqrt{n}} \sum_{i \in \II_1}  \triangle_{i}  \tau_2(\V_i)}_{=:T_3}.
\end{align*}
To control the first term $T_1$, condition on $(D_2^{\rct},\dataobs)$. Then $\tau_2$ is deterministic and $\{\psi_i \tau_2(\V_i)\}_{i\in\II_1}$ are conditionally independent and centered under the null hypothesis in Equation~\eqref{eq:nullg}. Thus, it suffices to show that the variance goes to zero:
\begin{align*}
\var[T_1\mid \dataobs]
&= 
\EE\left[\EE\left[T_1^2\mid D_2^{\rct},\dataobs\right]\mid \dataobs\right]
\notag\\
&=
\EE\!\left[\frac{1}{n}\sum_{i\in \II_1}\EE\left[\psi_i^2\tau_2(\V_i)^2\mid D_2^{\rct},\dataobs\right]\middle|\dataobs\right]
\notag\\
&=
\EE\!\left[\frac{1}{n}\sum_{i\in \II_1}\EE\!\left[\EE\left[\psi_i^2\mid \V_i\right]\tau_2(\V_i)^2\middle| D_2^{\rct},\dataobs\right]\middle|\dataobs\right]
\notag\\
&\lesssim
\EE\!\left[\frac{1}{n}\sum_{i\in \II_1}\tau_2(\V_i)^2\middle|\dataobs\right]
=
\EE\left[\tau_2(\V)^2\mid \dataobs\right]
=
o_{\pobs}(1).
\end{align*}
where we used that the conditional second moment $\EE\left[\testrv_{\trueg}^2\mid \V\right]$ is uniformly bounded, since the outcome $Y$ and the tolerance function $\estimandobs_\pm$ are both bounded, together with~\Cref{eq:tau2_bound}. Hence, 
by Chebyshev and iterated expectation, $|T_1|=o_{\mathbb P}(1)$. By symmetry, the same argument with $(D_1^{\rct},\dataobs)$ shows that $|T_2|=o_{\mathbb P}(1)$.\\\\
For $T_3$, we use Markov's inequality, so it is enough to show that the conditional first moment is $o_{\pobs}(1)$. By the triangle inequality and linearity of conditional expectation,
\begin{align*}
\EE\left[\left|T_3\right| \mid \dataobs\right]
&\le
\frac{1}{\sqrt n}\sum_{i\in \II_1}\EE\left[\left|\triangle_i\tau_2(\V_i)\right| \mid \dataobs\right]
\\
&=
\sqrt{n}\,\EE\left[\left|\triangle\,\tau_2(\V)\right| \mid \dataobs\right]
\\
&\le
\sqrt{n}\,\sqrt{\EE\left[\triangle^2 \mid \dataobs\right]}\sqrt{\EE\left[\tau_2(\V)^2 \mid \dataobs\right]}
=o_{\pobs}(1).
\end{align*}
where the last step uses~\Cref{eq:tau2_bound}. By Markov's inequality and iterated expectations, this yields $|T_3|=o_{\mathbb P}(1)$. Therefore, by the triangle inequality,
\begin{equation*}
\left|n\tstathat^2(\hat \psi_{\trueg})-n\tstathat^2(\psi_{\trueg})\right|=o_{\mathbb P}(1).
\end{equation*}

\paragraph{Step 2: controlling $\varhat{\hat{\psi}_{\trueg}}$}
As a second step, we control the variance term $\varhat{\hat{\psi}_{\trueg}}$. Our goal is again to show that 
\begin{equation*}
    \left\vert n\varhat{\hat \psi_{\trueg}}- n\varhat{ \psi_{\trueg}} \right\vert = o_{\mathbb P}(1).
\end{equation*}
Given the results from the previous paragraph in Equation~\eqref{eq:control_of_mean}, we note that it suffices to show that 
\begin{equation*}
    \left\vert  \frac{1}{n} \sum_{i \in \II_1} \hat \psi_{i}^2 \left(\frac{1}{\sqrt{n}} \sum_{j \in \mathcal I_2} k(\V_i, \V_j) \hat  \psi_j\right)^2  -  \frac{1}{n} \sum_{i \in \II_1} \psi_{i}^2 \left(\frac{1}{\sqrt{n}} \sum_{j \in \mathcal I_2} k(\V_i, \V_j)  \psi_j\right)^2 \right\vert = o_{\mathbb P}(1).
\end{equation*}
We  begin by decomposing the difference of the two terms on the LHS into the following six terms:
\begin{align}
    &= \underbrace{\frac{1}{n} \sum_{i \in \II_1} \triangle_{i}^2
    \left(\tfrac{1}{\sqrt{n}} \textstyle\sum_{j \in \mathcal I_2} k(\V_i, \V_j) (\psi_{j} + \triangle_{j})\right)^2}_{=:T_1}
    + \underbrace{\frac{1}{n} \sum_{i \in \II_1} (\psi_{i} + \triangle_{i})^2 \tau_2(\V_i)^2}_{=:T_2}
    \nonumber\\
    &\quad - \underbrace{\frac{1}{n} \sum_{i \in \II_1} \triangle_{i}^2 \tau_2(\V_i)^2}_{=:T_3}
    + \underbrace{\frac{2}{n} \sum_{i \in \II_1} \psi_{i}^2
    \left(\tfrac{1}{\sqrt{n}} \textstyle\sum_{j \in \mathcal I_2} k(\V_i, \V_j) \psi_{j}\right) \tau_2(\V_i)}_{=:T_4}
    \nonumber\\
    &\quad + \underbrace{\frac{4}{n} \sum_{i \in \II_1} \psi_{i} \triangle_{i}
    \left(\tfrac{1}{\sqrt{n}} \textstyle\sum_{j \in \mathcal I_2} k(\V_i, \V_j) \psi_{j}\right) \tau_2(\V_i)}_{=:T_5}
    \nonumber\\
    &\quad + \underbrace{\frac{2}{n} \sum_{i \in \II_1} \psi_{i} \triangle_{i}
    \left(\tfrac{1}{\sqrt{n}} \textstyle\sum_{j \in \mathcal I_2} k(\V_i, \V_j) \psi_{j}\right)^2}_{=:T_6}. \nonumber
\end{align}
We show that $\EE\left[|T_\ell|\mid \dataobs\right]=o_{\pobs}(1)$ for every $\ell\in\{1,\dots,6\}$. Then Markov's inequality together with iterated expectation implies $|T_\ell|=o_{\mathbb P}(1)$ for every $\ell\in\{1,\dots,6\}$.\\\\ 
\textit{Controlling $T_1$:} Since $T_1\ge 0$, it suffices to show that $\EE\left[T_1\mid \dataobs\right]=o_{\pobs}(1)$. Since
\[
\frac{1}{\sqrt{n}}\sum_{j\in \II_2}k(\V_i,\V_j)(\psi_j+\triangle_j)
=
\bar\tau_2(\V_i)+\tau_2(\V_i),
\]
and using $(a+b)^2\le 2a^2+2b^2$, we have
\[
T_1
\le
\frac{2}{n}\sum_{i\in \II_1}\triangle_i^2\bar\tau_2(\V_i)^2
+
\frac{2}{n}\sum_{i\in \II_1}\triangle_i^2\tau_2(\V_i)^2.
\]
We can then bound each term separately. We have
\begin{align*}
\EE\!\left[\frac{1}{n}\sum_{i\in \II_1}\triangle_i^2\bar\tau_2(\V_i)^2 \,\middle|\, \dataobs\right]
=
\EE\left[\triangle^2\bar\tau_2(\V)^2\mid \dataobs\right]
=
o_{\pobs}(1),
\end{align*}
where we used~\Cref{eq:black_delta_bartau2_bound}. 
Moreover,
\begin{equation}
\label{eq:black_delta_tau2_bound}
\begin{aligned}
\EE\!\left[\frac{1}{n}\sum_{i\in \II_1}\triangle_i^2\tau_2(\V_i)^2 \,\middle|\, \dataobs\right]
&=
\EE\left[\triangle^2\tau_2(\V)^2\mid \dataobs\right]
\\
&\lesssim
\EE\!\left[\triangle^2\sum_{j\in \II_2}\triangle_j^2 \,\middle|\, \dataobs\right]
\\
&=
\EE\left[\triangle^2\mid \dataobs\right]\,
\EE\!\left[\sum_{j\in \II_2}\triangle_j^2 \,\middle|\, \dataobs\right]
\\&=
o_{\pobs}(1).
\end{aligned}
\end{equation}
Here we used boundedness of the kernel, Cauchy-Schwarz, and that the generic pair $(\V,\triangle)$ is conditionally independent of the samples from the second fold given $\dataobs$.\\\\
\textit{Controlling $T_2$ and $T_3$:} We can again upper-bound the expectation and apply Markov's inequality:
\begin{align*}
\EE\left[|T_2|\mid \dataobs\right]
&\lesssim
\EE\left[\tau_2(\V)^2\mid \dataobs\right]
+
\EE\left[\triangle^2\tau_2(\V)^2\mid \dataobs\right]
=
o_{\pobs}(1),
\end{align*}
using that $(\psi+\triangle)^2\le 2\psi^2+2\triangle^2$ and combining the bounds in~\Cref{eq:tau2_bound,eq:black_delta_tau2_bound}. 
 The same argument applies by symmetry to $T_3$. \\\\
\textit{Controlling $T_4$:} By Cauchy-Schwarz and~\Cref{eq:tau2_bound,eq:bartau2_conditional},
\begin{equation*}
\begin{aligned}
\EE\left[|T_4|\mid \dataobs\right]
&\lesssim
\EE\left[|\bar\tau_2(\V)\tau_2(\V)|\mid \dataobs\right]
\\
&\le
\sqrt{\EE\left[\bar\tau_2(\V)^2\mid \dataobs\right]}\sqrt{\EE\left[\tau_2(\V)^2\mid \dataobs\right]}
=
o_{\pobs}(1).
\end{aligned}
\end{equation*}
\textit{Controlling $T_5$:} By Cauchy-Schwarz,
\begin{equation*}
\begin{aligned}
\EE\left[|T_5|\mid \dataobs\right]
&\lesssim
\EE\left[|\triangle\,\bar\tau_2(\V)\tau_2(\V)|\mid \dataobs\right]
\\
&\le
\sqrt{\EE\left[\triangle^2\bar\tau_2(\V)^2\mid \dataobs\right]}
\sqrt{\EE\left[\tau_2(\V)^2\mid \dataobs\right]}
\\
&\lesssim
r_n\sqrt{\EE\left[\tau_2(\V)^2\mid \dataobs\right]}
=
o_{\pobs}(1),
\end{aligned}
\end{equation*}
by~\Cref{eq:black_delta_bartau2_bound,eq:tau2_bound}.

\textit{Controlling $T_6$:} By Cauchy--Schwarz,
\begin{equation*}
\begin{aligned}
\EE\left[|T_6|\mid \dataobs\right]
&\lesssim
\EE\left[|\triangle|\,\bar\tau_2(\V)^2\mid \dataobs\right]
\\
&\le
\sqrt{\EE\left[\triangle^2\bar\tau_2(\V)^2\mid \dataobs\right]}
\sqrt{\EE\left[\bar\tau_2(\V)^2\mid \dataobs\right]}
\\
&\lesssim
r_n
=
o_{\pobs}(1),
\end{aligned}
\end{equation*}
where we used~\Cref{eq:black_delta_bartau2_bound,eq:bartau2_conditional}. Collecting the bounds above, Markov's inequality together with iterated expectations yields $|T_\ell|=o_{\mathbb P}(1)$ for every $\ell\in\{1,\dots,6\}$. Therefore, by the triangle inequality,
\begin{equation*}
\left|n\varhat{\hat\psi_{\trueg}}-n\varhat{\psi_{\trueg}}\right|
\le
\sum_{\ell=1}^6 |T_\ell|
=
o_{\mathbb P}(1).
\end{equation*}

\subsection{Power of the test}
\label{sec:power}
We discuss our result on the power of the test in rejecting the alternative hypothesis. While \citet{hussain2023falsification,muandet2020kernel} show asymptotic normality for the conditional moment test statistics $\mathbb M^2$ under the alternative hypothesis that $\mathbb M^2 \neq 0$, the same result does not directly apply to our test statistic. However, as we show in the following theorem, our test statistic grows at a rate $\sqrt{n}$. For the sake of clarity, we only prove the result for the oracle test statistic $\tstatopt$ computed from $\testrv_g$. Nevertheless, we remark that our result can be extended to the empirical test statistic $\tstathatopt$ via the same arguments used in the proof of Theorem~\ref{thm:main}. 
\begin{thm}
\label{thm:power}
Assume that for every $\epsilon>0$, the function class $\mathcal{G}$ has a finite $\ell_{\infty}-$norm covering number. Then, we can lower-bound the 
test statistic, in probability as $n \to \infty$, by \begin{equation*}
    \tstatopt = \underset{\g \in \GG}{\min}\left| \frac{\sqrt{n}~\tstathat^2( \psi_{g})}{\hat \sigma \left( \tstathat^2( \psi_g) \right)  } \right\vert \gtrsim \sqrt{n}~ \left(\inf_{g \in \mathcal G} \sup _{\|f\|_\FF \leq 1}\EE\left[\psi_{g}(Z) f(\V)\right]\right)^2,
\end{equation*}
where we use $\gtrsim$ to hide universal constants not depending on $n$.
\end{thm}
Thus, under the alternative hypothesis 
\begin{equation*}
    H_A^{\GG }: \inf_{g \in \GG} \sup _{\|f\|_\FF \leq 1}\EE\left[\psi_{g}(Z) f(\V)\right]  >0,
\end{equation*}
 the RHS grows at a rate $\sqrt{n}$, which implies that our test has an asymptotic power of one~(note that the same rate is achieved by existing conditional moment tests~\citep{hussain2023falsification,muandet2020kernel}). 
 \paragraph{Proof of Theorem~\ref{thm:power}}
Let us define $$T:= \inf_{g \in \mathcal G} \sup _{\|f\|_\FF \leq 1}\EE\left[\psi_{g}(Z) f(\V)\right],$$ and note that if $T =0$ the result follows trivially. Thus, we may assume that $T>0$ is some constant independent of $n$. Additionally, observe that $\testrv_g$ is uniformly bounded since the outcome $Y$ is a bounded random variable. Therefore,  the variance term $\hat \sigma \left( \tstathat^2( \psi_g) \right)$ is also uniformly bounded, and it suffices to show that 
$\inf_{g \in \mathcal G}\tstathat^2( \psi_{g}) = \Omega_{\mathbb P}(1)$ is lower bounded in probability.

\paragraph{Controlling $\tstathat^2( \psi_{g})$}
First, recall that our test statistic is given by 
\begin{equation}
\label{eq:tstatapx}	
\tstathat^2( \testrv_{g}) =   \frac{1}{n^2} \sum_{i=1}^{n}  \sum_{j=n +1}^{2n} \testrv_g(Z_i) k(\subx_i, \subx_j) \testrv_g(Z_j).
\end{equation}
Further, for all $g \in \mathcal G$, it holds that 
\begin{align*}
    \EE\left[\tstathat^2( \testrv_{g})\right] = \EE\left[\testrv_g(Z) k(\V, \tilde{X}^\J) \testrv_g(\tilde Z)\right] \geq \inf_{g \in \mathcal G} \mathbb \EE\left[\testrv_g(Z) k(\V, \tilde{X}^\J)\testrv_g(\tilde Z)\right]= T^2,
\end{align*}
where $\tilde Z$ is an independent copy of $Z$ following the same distribution, and the last equality follows from Equation~\eqref{eq:defH}. Thus, it suffices to show that the following inequality holds with probability tending to one as $n\to \infty$
\begin{equation*}
\label{eq:proofA1}
\sup_{g\in \mathcal G} \left \vert \tstathat^2( \testrv_{g}) - \EE\left[\tstathat^2( \testrv_{g})  \right] \right \vert \leq \frac{T^2}{2}.
\end{equation*}

We use a simple $\epsilon$-net argument to show this result. Let $\mathcal G_{\epsilon}$ be the epsilon net in $\ell_{\infty}$ distance of balls with radii $\epsilon$. 
Then, since $\testrv_g$ is uniformly bounded, it holds that for all $Z$ and $g \in\mathcal G$, \begin{equation*}
\inf_{\tilde g \in \GG_\epsilon}\vert\psi_g(Z) - \psi_{\tilde g}(Z)\vert \lesssim \epsilon.
\end{equation*}
Thus, from the definition of $\tstathat^2(\psi_g)$ in Equation~\eqref{eq:tstatapx} it follows that we can choose a constant $\epsilon>0$, such that
the following inequality holds almost surely,
\begin{equation}
\label{eq:boundgtilde}
      \sup_{g \in \mathcal G}~ \inf_{\tilde{g} \in \mathcal G_{\epsilon} }~
    \left \vert \tstathat^2( \psi_{g}) - \tstathat^2( \psi_{\tilde{g}})\right \vert \leq \frac{T^2}{8}.
\end{equation}
By the same argument used to establish~\Cref{eq:boundgtilde}, we also obtain
\[
\sup_{g \in \mathcal G}~ \inf_{\tilde{g} \in \mathcal G_{\epsilon} }~
    \left \vert \EE\left[\tstathat^2( \psi_{g})\right] - \EE\left[\tstathat^2( \psi_{\tilde{g}})\right]\right \vert \leq \frac{T^2}{8}.
\]
Hence, by the triangle inequality,
\begin{equation}
\label{eq:boundgtilde_centered}
\sup_{g \in \mathcal G}~ \inf_{\tilde{g} \in \mathcal G_{\epsilon} }~
\left \vert \left(\tstathat^2( \psi_{g}) - \EE\left[\tstathat^2( \psi_{g})\right]\right) - \left(\tstathat^2( \psi_{\tilde{g}}) - \EE\left[\tstathat^2( \psi_{\tilde{g}})\right]\right)\right \vert \leq \frac{T^2}{4}.
\end{equation}
Then, it holds that \begin{align*}
    &~\mathbb P \left (~\sup_{g \in \GG} \left | \tstathat^2(\psi_g) - \EE\left[\tstathat^2(\psi_g)\right]\right | \leq \frac{T^2}{2} \right ) \\
   & \overset{(i)}{\geq}   \mathbb P \left(  ~\sup_{\tilde g \in \mathcal G_{\epsilon}}\left | \tstathat^2( \psi_{\tilde g}) - \EE\left[\tstathat^2( \psi_{\tilde g}) \right] \right |  \leq \frac{T^2}{4}\right ) \\
    &\overset{(ii)}{\geq} 1 - \sum_{\tilde g \in \GG_\epsilon} \underbrace{\mathbb P \left ( \left | \tstathat^2( \psi_{\tilde g}) - \EE\left[\tstathat^2( \psi_{\tilde g}) \right] \right |\geq \frac{T^2}{4} \right )}_{\overset{n \to \infty}{\to}~0 \;\; \text{(L.L.N.)}} , 
\end{align*}
where (i) follows from applying the inequality in~\Cref{eq:boundgtilde_centered} and (ii) follows since, by assumption, for every fixed $\epsilon>0$ the cover $\vert \mathcal G_{\epsilon}\vert <\infty$ is constant as a function of $n$.

\clearpage
\section{Additional experiments}

\subsection{Ablation study of the  feature subset $\subx$}
\label{sec:ablation_subset}
In Scenario 2 from~\Cref{fig:heatmap_scenario2}, we introduced constant bias within each subgroup resulting from different combinations of the features \texttt{newbie}, \texttt{mens} and \texttt{channel}, with a maximum true bias $\truedelta=60$. Figure~\ref{fig:ablation_v} shows the effect of the selected feature set $\subx$ on the average lower bound $\deltalb$ for the bias model from Scenario 2. When $|\J|=3$, we select the features that capture the bias between rct and os datasets~(\texttt{newbie}, \texttt{mens}, \texttt{channel}), and hence we achieve the highest power. Intuitively, if the feature set is smaller, some of the bias averages out, and the test loses power. On the other hand, when increasing the feature set, the test loses power due to the curse of dimensionality, being particularly severe with smaller sample sizes.  After $|\J|=6$ (i.e. $\subx=X$), we add redundant features sampled from a standard normal distribution $\mathcal{N}(0, 1)$.

\subsection{Convergence of the optimization procedure}
\label{apx:opt}
\begin{figure*}[t]
    \centering
      \begin{subfigure}[b]{0.48\textwidth}
        \centering
\includegraphics[width=0.55\textwidth]{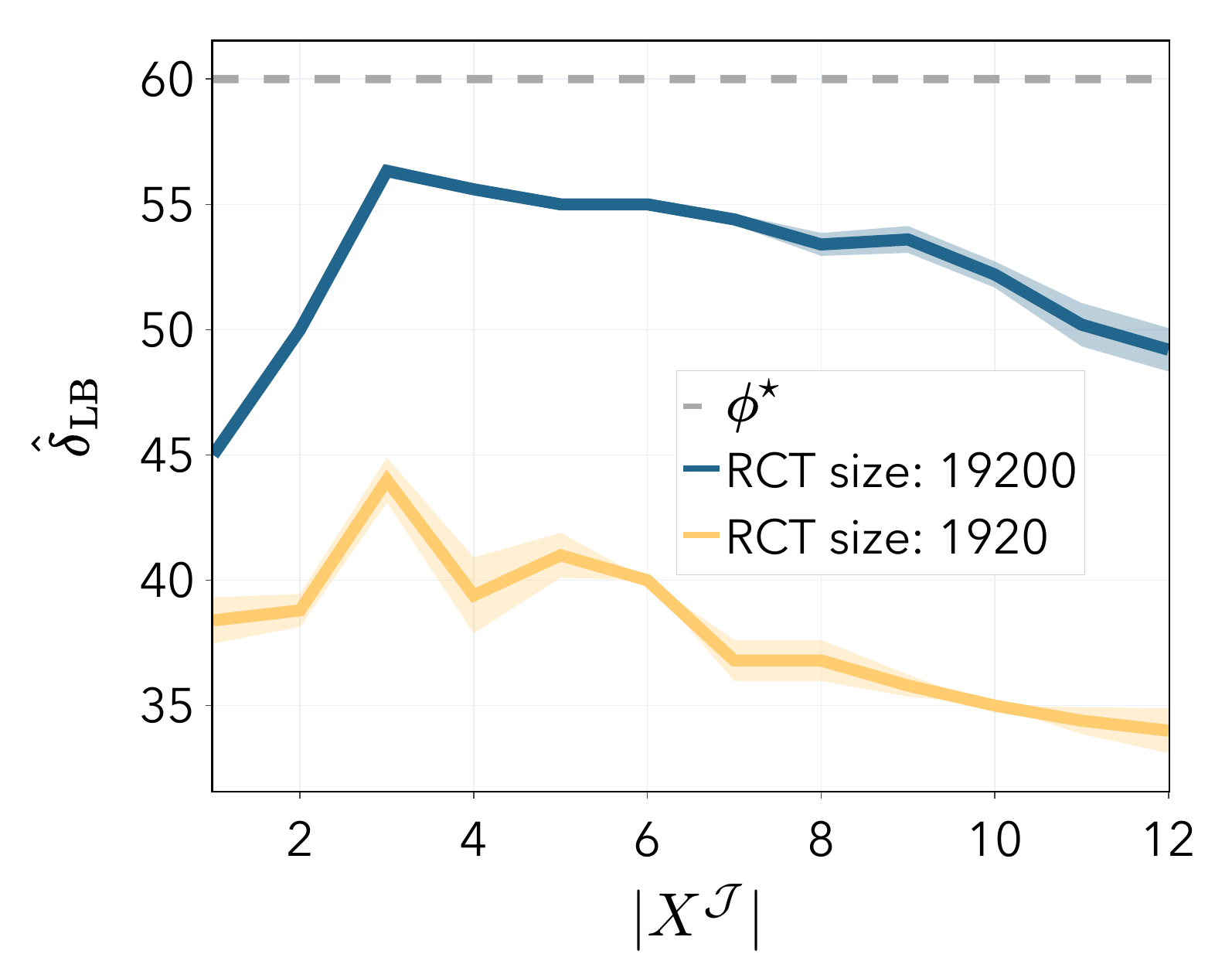}
        \caption{Ablation over subset of features
        }
  \label{fig:ablation_v}
    \end{subfigure}
    \hfill
    \begin{subfigure}[b]{0.48\textwidth}
        \centering
\includegraphics[width=0.6\textwidth]{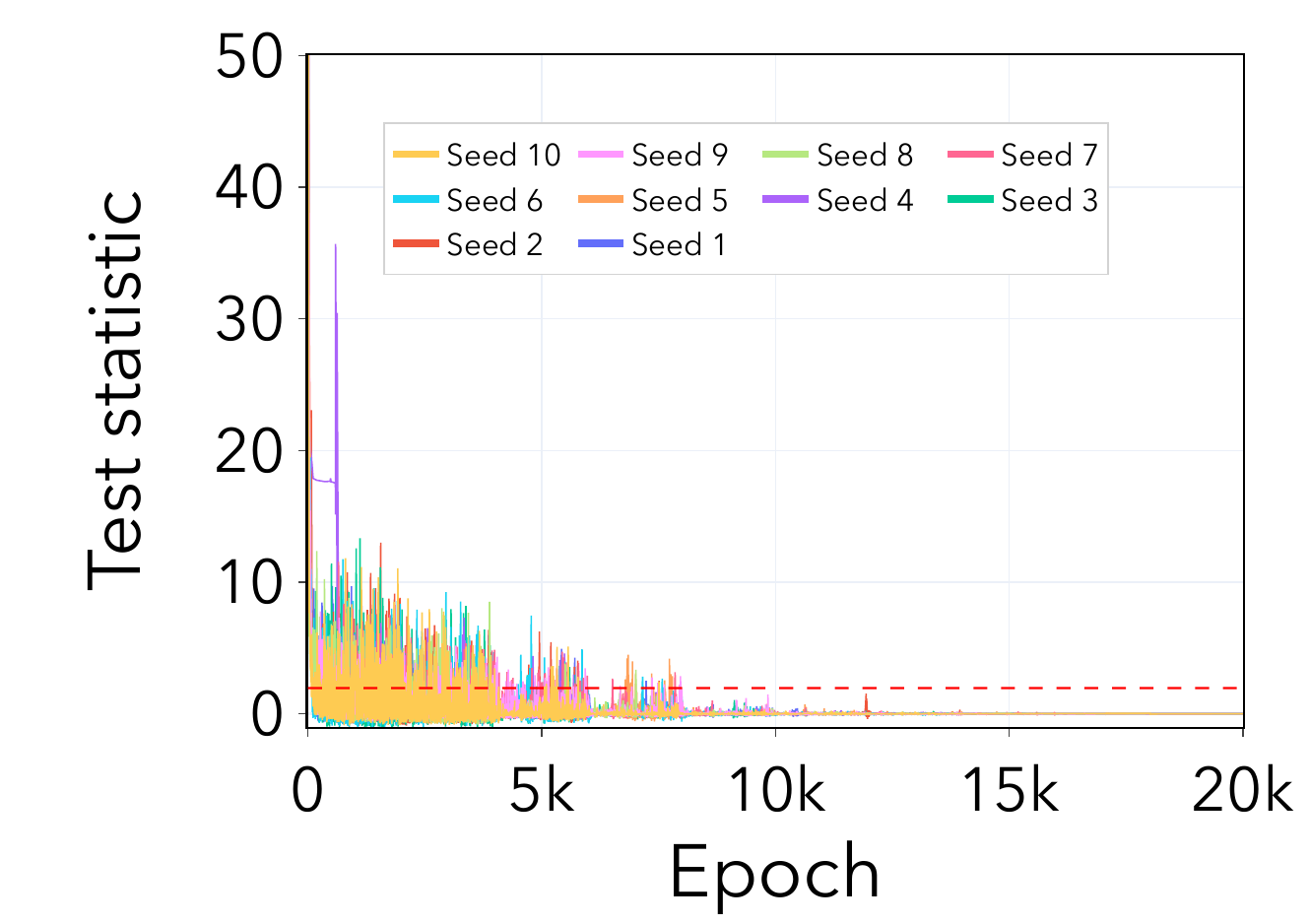}
      \caption{Optimization convergence}
 \label{fig:opt}
    \end{subfigure}
    \caption{For all the plots: the significance level is set at $\alpha=0.05$, and the bias model is from Scenario 2. (a) Effect of varying the feature set $\subx$ on the average lower bound $\deltalb$, illustrating the trade-off between feature set size and the power of the test. $\phi^\star$ represents the oracle test, which rejects for $\delta<\truedelta$. The highest power is achieved when the feature set size $|\J|=3$, including only the relevant features to model the bias.  We average runs over 5 seeds and report the standard error. (b) Evolution of the test statistic with respect to the training epochs using the small neural network. We set the user tolerance to $\delta=58$, close to the maximum true bias $\truedelta =60$. The dashed red line represents the $(1-\alpha)$-quantile of the absolute standard normal distribution. The rest of the hyperparameters are the same as in the experimental setting from \Cref{sec:exp}.}
\end{figure*}

We provide evidence that our testing procedure is reliable,  meaning that the optimizer consistently reaches the same solution for the bias model from Scenario 2 and the small neural network model. Recall that, given the non-convex nature of the optimization problem, we cannot guarantee convergence to the true global minimizer $\trueg$. \Cref{fig:opt} shows the test statistic as a function of the training epoch under different random network initializations. We observe that the test statistic consistently reaches the same minimum and that the optimization stabilizes after $10000$ epochs.

\subsection{Interpretability of the testing procedure}

Similar to the test proposed by \citet{hussain2023falsification}, our testing procedure outputs a ``witness function'' that enables practitioners to identify the most biased subgroups within the observational dataset. Additionally, our witness function provides insights into the bias strength and direction for each subgroup. This is achieved by minimizing the objective in \Cref{eq:test}, where we learn the bias function $\hat g$. If the function class $\GG$ is sufficiently rich to model the bias structure, and the optimizer converges to a global minimizer $\trueg$, we expect $\hat g$ to be a good approximation of $\trueg$.

\begin{figure}[t]
    \centering
    \begin{subfigure}{0.35\textwidth}
        \centering
        \includegraphics[width=\linewidth]{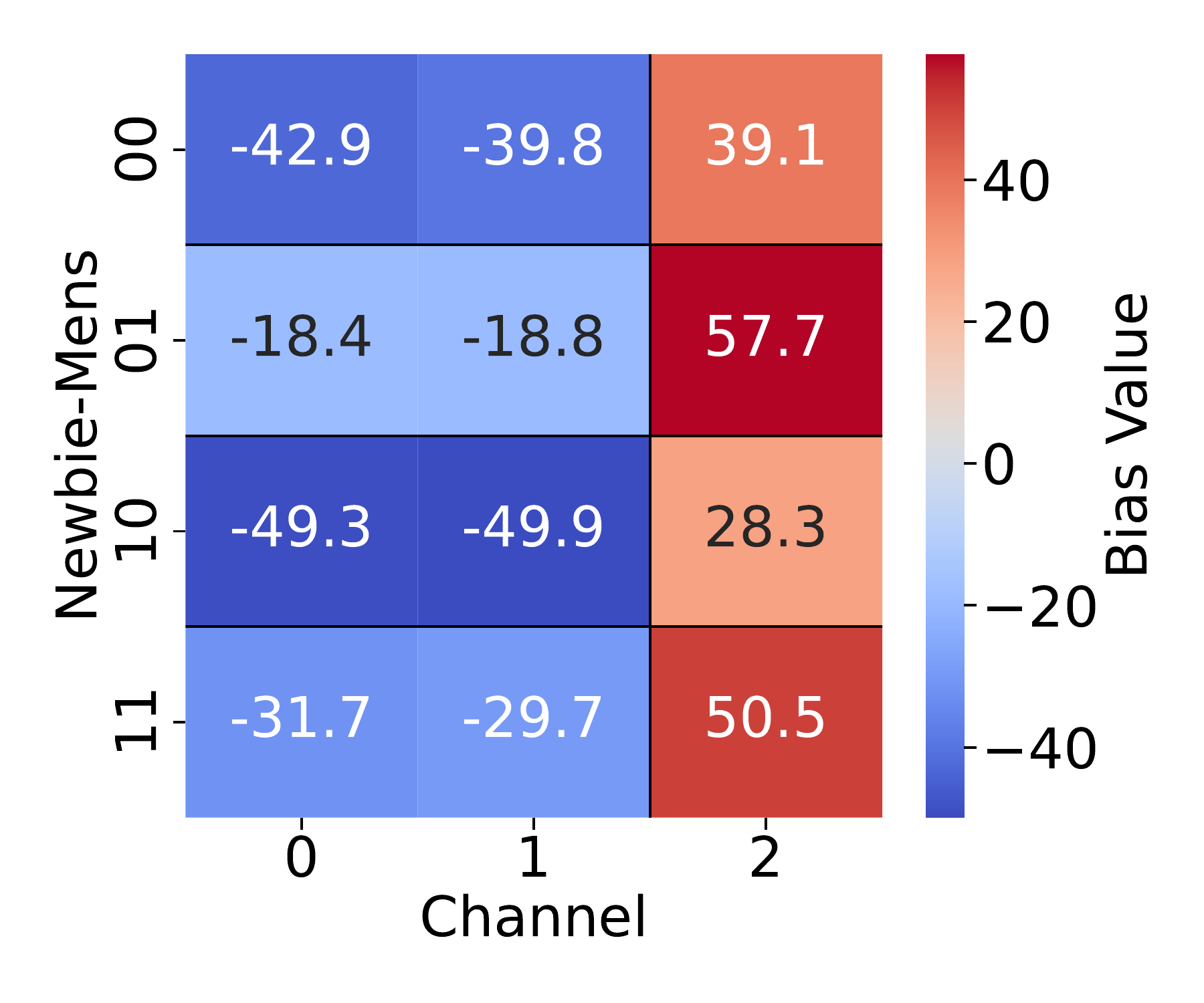}
        \caption{Estimated bias}
    \end{subfigure} 
    \begin{subfigure}{0.35\textwidth}
        \centering
        \includegraphics[width=\linewidth]{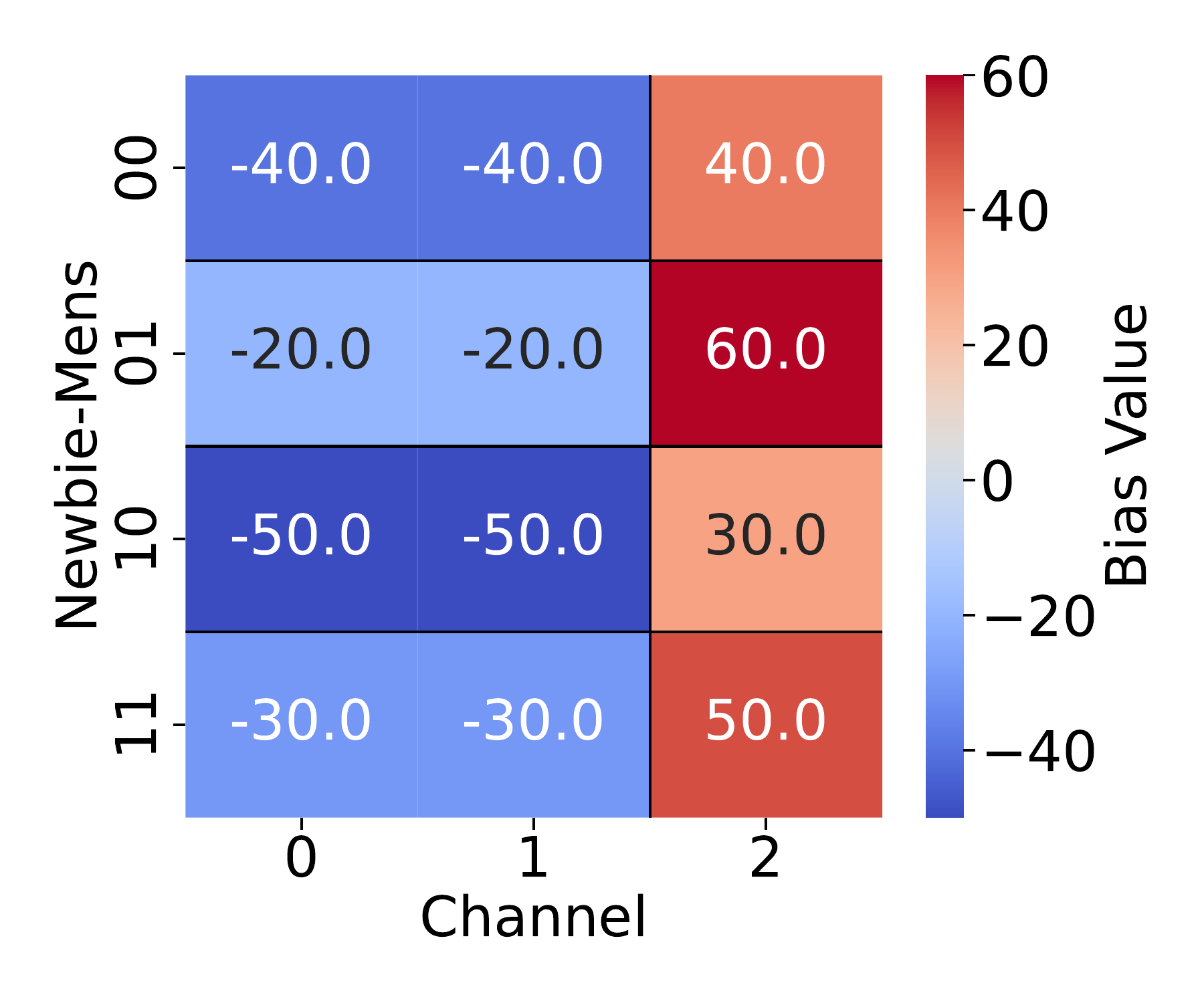}
        \caption{True bias model}
    \end{subfigure}
\caption{Comparison between the estimated and true bias models for Scenario 2. Our estimates of the bias from~\Cref{eq:groupbias} closely align with the true bias. We run the test with a random seed, using the same hyperparameters as in our experimental evaluation, and set the user tolerance to $\delta=57$.}
    \label{fig:interpretability}
\end{figure}

To interpret this bias function, we observe that $\hat g(X) \in [0, 1]$ interpolates between the tolerance bounds $\lbobs(X)$ and $\ubobs(X)$; therefore, values close to zero indicate a negative bias of magnitude close to user-tolerance $\delta$, while values close to one indicate the same for positive bias. Hence, we can estimate the subgroup bias as \begin{align}
\label{eq:groupbias} \texttt{bias}(G) =\deltalb \left(\frac{2}{|G|}\sum_{X_i \in G}\hat g\left(X_i\right)-1\right), \end{align} where $G$ represents the subgroup of interest. In~\Cref{fig:interpretability}, we illustrate how practitioners could use the witness function for Scenario 2, where the categorical nature of the features defines subgroups. We compare the estimated bias with the ground truth and observe that our estimates closely align with the true bias model. In scenarios where subgroups are not predefined, a practitioner can select the bottom or top 10\% of witness function values, as suggested by \citet{hussain2023falsification}.

\clearpage
\section{Experimental details}

\subsection{Hillstrom's MineThatData}

Hillstrom's MineThatData Email dataset \citep{hillstrom2008} is a large-scale, real-world randomized trial that contains records of 64,000 customers who made purchases online within the last twelve months. They were part of an email campaign designed to assess the effectiveness of different campaign strategies. Two treatment groups, ``Men’s'' and ``Women’s'' email campaigns, and a control group were established, with treatments randomly assigned. Our analysis primarily focuses on a combined treatment group, which constitutes approximately 66\% of the dataset. Although the original dataset presents various outcomes, including binary indicators of customers visiting or purchasing in the days after the campaign, we focus on the dollars spent in the two weeks post-campaign. The dataset provides data on annual spending (\texttt{history}), merchandise type (\texttt{mens} and \texttt{womens}), geographical location (\texttt{zip code}), newcomer status (\texttt{newbie}), and purchasing avenues (\texttt{channel}). We, therefore, discard features describing the history segment (\texttt{history segment}) and recency of the last purchase (\texttt{recency}). Since the average treatment effect is close to zero, we add a constant shift of 30 to all treated individuals, allowing us more flexibility to introduce bias. We normalize continuous features and one-hot-encode categorical features, resulting in a 13-dimensional dataset. By default, we use 80\% of the full dataset as the observational study (os), and the remaining 20\% as the randomized controlled trial (rct).

We fit the propensity score using logistic regression with default hyperparameters from \texttt{scikit-learn}. We train a \texttt{Random Forest Classifier} for the selection score (rct or os), also with default hyperparameters from \texttt{scikit-learn}. Finally, we estimate the CATE functions using the doubly-robust learner from \citet{kennedy2023towards}, instantiating \texttt{Random Forest Regressors} for the potential outcome functions and the pseudo-outcome regression, fixing hyperparameters to 300 \texttt{tree estimators} with a \texttt{maximum depth} of 6.

\begin{figure}[ht]
    \centering
    \begin{subfigure}{0.37\textwidth}
        \centering
        \includegraphics[width=\linewidth]{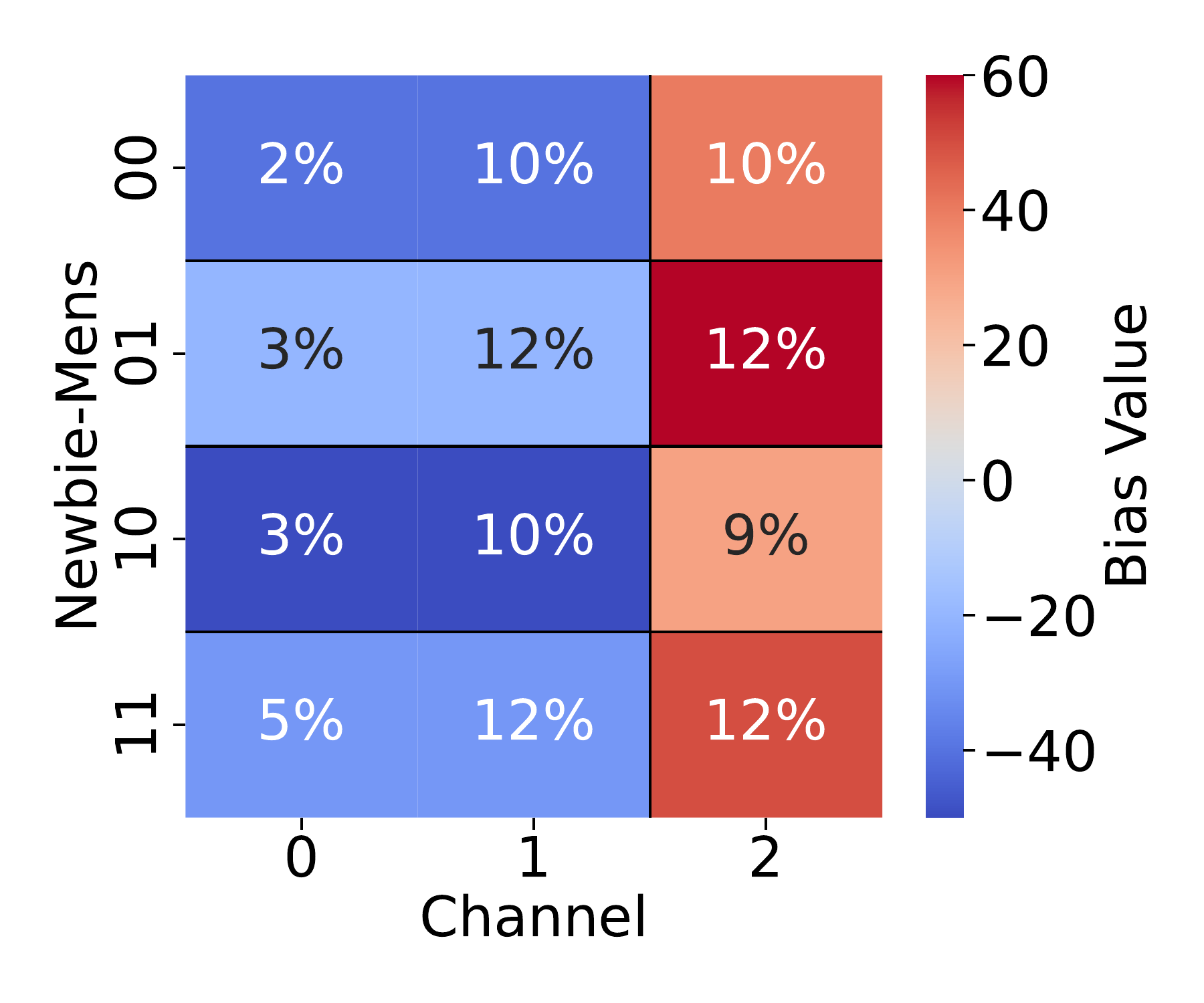}
        \caption{Scenario 2}
        \label{fig:heatmap_scenario2}
    \end{subfigure} 
    \begin{subfigure}{0.37\textwidth}
        \centering
        \includegraphics[width=\linewidth]{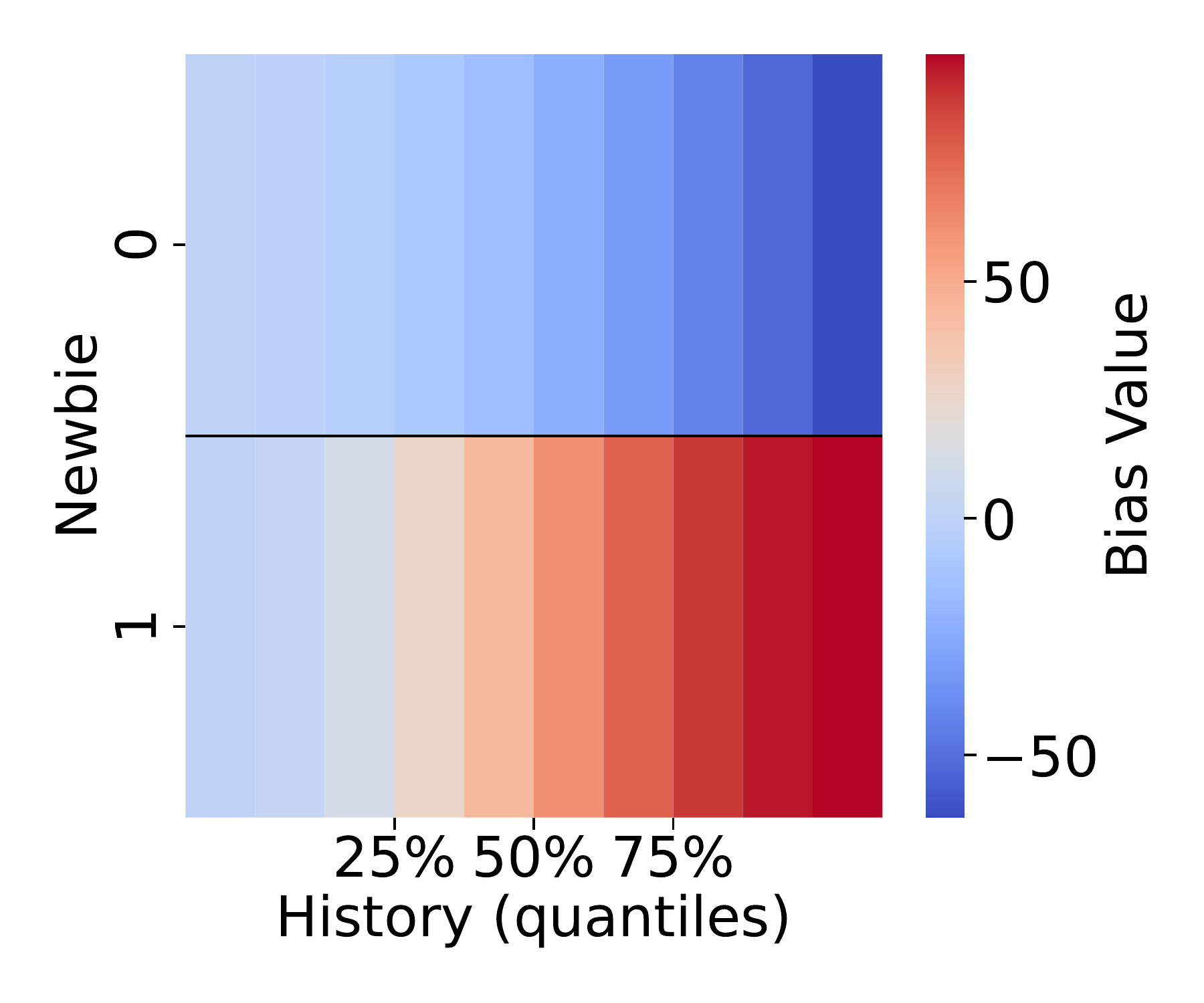}
        \caption{Scenario 3}
        \label{fig:heatmap_scenario3}
    \end{subfigure}
\caption{Heatmap visualizations of the bias for (a) Scenario 2 based on 12 subgroups with different biases (the numbers in the cells represent the percentage w.r.t. the full observational dataset), and (b) Scenario 3 based on a quadratic polynomial bias.} 
    \label{fig:scenarios}

\end{figure}

\paragraph{Bias models} We illustrate the bias model for Scenario 2 and Scenario 3 in~\Cref{fig:scenarios}. For scenario 3, we sample the coefficient for the polynomial bias model in~\Cref{fig:heatmap_scenario3} from a normal distribution  $\mathcal{N}(0, 0.01^2)$.

\subsection{Women's Health Initiative}
\label{apx:whi_exp}
The Women's Health Initiative (WHI) is a long-term national health study that has focused on strategies for preventing the major causes of death, disability, and frailty in older women, specifically heart disease, cancer, and osteoporotic fractures. This multi-million dollar, 20+ year project, sponsored by the National Institutes of Health (NIH) and the National Heart, Lung, and Blood Institute (NHLBI), initially enrolled 161,808 women aged 50-79 between 1993 and 1998. The WHI was one of the most definitive, far-reaching clinical trials of post-menopausal women's health ever undertaken in the US.  

The WHI had two major parts: a randomized trial and an observational study. The randomized trial enrolled 68,132 women in trials testing three prevention strategies. Eligible women could choose to enroll in one, two, or three of the trial components.
\begin{itemize}
\item A Hormone Therapy Trial (HT) that examined the effects of combined hormones or estrogen alone on the prevention of heart disease and osteoporotic fractures and associated risk for breast cancer. 
\item 	A Dietary Modification Trial (DM) that evaluated the effect of a low-fat and high-fruit, vegetable, and grain diet on preventing breast and colorectal cancers and heart disease. \item A Calcium and Vitamin D Trial (CaD) that evaluated the effect of calcium and vitamin D supplementation on preventing osteoporotic fractures and colorectal cancer. \end{itemize}
The Observational Study (OS) examines the relationship between lifestyle, health risk factors, and disease outcomes. This component involves tracking the medical events and health habits of 93,676 women. Recruitment for the observational study was completed in 1998, and participants have been followed since.

We use observational study and randomized trial data from the Women’s Health Initiative (WHI) to assess our method in a real-world scenario. We use the Hormone Therapy (HT) trial as the RCT in our analysis ($\nrct=16,608$), run on postmenopausal women aged 50-79 years with an intact uterus. The trial investigated the effect of hormone therapy on several types of cancers, cardiovascular events, and fractures, measuring the “time-to-event” for each outcome. In the WHI setup, the observational study component was run in parallel, and outcomes were tracked similarly to those of the RCT. 

\vspace{-1mm}
\paragraph{Data preprocessing} 
We binarize the outcome, where $Y = 1$ if coronary heart disease was observed in the first seven years of follow-up, and $Y = 0$ otherwise. To establish treatment and control groups in the observational study, we use questionnaire data in which participants confirm or deny usage of combination hormones (i.e. both estrogen and progesterone) in the first three years. Using this procedure, we end up
 with a total of $\nobs = 33,511$ patients. Finally, we restrict the set of covariates used to those that are measured in both the RCT and the observational study.  In particular, we use as covariates only those measured in both the RCT and observational study, and we further restrict them to those identified as significant in epidemiological literature, such as in~\citep{prentice2005combined}. Specifically, the covariates in our analysis are: \texttt{AGE}, \texttt{ETHNIC\_White}, \texttt{BMI}, \texttt{SMOKING\_Past\_Smoker}, \texttt{SMOKING\_Current\_Smoker}, \texttt{EDUC\_x\_College\_graduate\_or\_Baccalaureate Degree}, \\\texttt{EDUC\_x\_Some\_post-graduate\_or\_professional}, \texttt{MENO}, \texttt{PHYSFUN}. The data used is available on \href{https://biolincc.nhlbi.nih.gov/studies/whi_ctos}{BIOLINCC}.

\paragraph{Experimental details} We use a gaussian kernel with $\texttt{bandwidth}=1.0$. The set of features for the granularity of the test is chosen to be $\J=\{\texttt{AGE},\texttt{MENO}\}$. We use a logistic regression model for both the outcome model and propensity score~(default hyperparameters in \texttt{scikit-learn} were used). We train a neural (1 hidden layer and 10 neurons) network with \texttt{Adam}, with a \texttt{learning rate} of $0.01$ for $500$ epochs. We repeat the optimization for $10$ seeds with different initializations to ensure that we converge. \vspace{-10mm}

\end{document}